\documentclass[sigconf, 9pt, balance=false]{acmart}

\usepackage{booktabs}
\usepackage{threeparttable}
\usepackage{subcaption}
\usepackage{algorithm}
\usepackage{algpseudocode}
\usepackage{multirow}
\usepackage{amsmath}
\usepackage{graphicx}
\usepackage{caption}
\usepackage{float}
\usepackage[normalem]{ulem}
\usepackage{enumitem}
\usepackage{hyperref}

\usepackage{xcolor}

\newlength{\colw}
\newlength{\colgap}

\newcommand{\colstack}[3]{%
  \begin{tabular}{@{}c@{}}
    \includegraphics[width=\colw]{#1}\\[-0.8mm]
    \includegraphics[width=\colw]{#2}\\[-0.8mm]
    \includegraphics[width=\colw]{#3}
  \end{tabular}%
}

\definecolor{revisionblue}{RGB}{0,0,255}

\useunder{\uline}{\ul}{}

\AtBeginDocument{%
  }

\setcopyright{none}
\setcopyright{none}
\copyrightyear{2027}
\acmYear{2027}
\acmDOI{}

\acmConference[SenSys '27]{ACM/IEEE International Conference on Embedded Artificial Intelligence and Sensing Systems}{2027}{New York, NY, USA}

\begin{document}

\title{Short Paper: Rendering the Field, Obeying the Physics: Scene-Aware 3D Gaussian Splatting for Wireless Radiation Fields}


%\author{Anonymous Author(s)}

\author{Jingzhou Shen}
\affiliation{%
  \institution{Florida International University}
  \city{Miami}
  \state{Florida}
  \country{USA}}
\email{jshen020@fiu.edu}  

\author{Xuyu Wang}
\authornote{Corresponding Author}
\affiliation{%
  \institution{Florida International University}
  \city{Miami}
  \state{Florida}
  \country{USA}}
\email{xuywang@fiu.edu}

\begin{abstract}
Reconstructing wireless radiation fields is a core capability for many sensing and communication systems. Recent neural-field and Gaussian-based representations perform well on this task, but they are mostly data-driven and ignore the physical laws that govern wave propagation. Their predictions can therefore become physically implausible where few measurements are available. Physics-informed neural networks~(PINNs) offer a way to address this, yet most existing approaches simply apply equation-based losses to network outputs, treating physics as a generic constraint that is blind to the propagation environment. We argue that effective physical supervision should be scene-aware, enforcing wave-propagation constraints in physically valid free-space regions while relaxing them near obstacles that violate these assumptions. To this end, we present PINN-GS, a 3D Gaussian Splatting (3DGS) framework that integrates physics into the rendering pipeline rather than appending it afterward. Our key idea is to jointly learn the underlying wave field and the propagation structure of the scene directly from measurements, so that physical constraints are applied selectively in free space and relaxed near obstacles, all without explicit geometry or manual scene labels. Experiments on real-world RSSI datasets and a public RFID spectrum benchmark show that PINN-GS consistently outperforms representative neural wireless reconstruction methods, improving reconstruction accuracy and producing more physically consistent field estimates. 
\end{abstract}
%%
%% The code below is generated by the tool at .


\copyrightyear{2027}\acmYear{2027}
\setcopyright{cc}
\setcctype{by}
\acmConference[SenSys '27]{ACM/IEEE International Conference on Embedded Artificial Intelligence and Sensing Systems}{May 17-20, 2027}{Boulder, CO, USA}
\acmBooktitle{ACM/IEEE International Conference on Embedded Artificial Intelligence and Sensing Systems (SenSys '27), May 17-20, 2027, Boulder, CO, USA}
\acmDOI{10.1145/3862575.3862840}
\acmISBN{979-8-4007-3124-2/2027/05}


\maketitle

\section{Introduction}

A wireless radiation field describes how electromagnetic energy emitted by a transmitter is distributed across a physical environment, accounting for propagation effects such as attenuation, multipath, and scattering. Reconstructing this field is a fundamental capability for wireless sensing and communication systems, where downstream applications often depend on the signal strength, channel state information, or spatial spectrum it induces~\cite{nerf2, wrfgs, 8424015}. As wireless deployments grow denser and integrated sensing and communication emerges as a key feature of next-generation networks~\cite{9737357}, there is a growing need for site-specific radiation fields that faithfully capture the geometry-dependent propagation structure of real environments. In practice, the field can only be probed at a limited number of transmitter or receiver locations~\cite{nerf2, 10.1145/3666025.3699363}, so it must be reconstructed from sparse in-situ measurements. Wireless radiation field reconstruction is therefore not only an electromagnetic propagation problem, but also a problem of recovering a structured physical field from incomplete observations.

Classical approaches to radiation field reconstruction typically rely on analytical propagation assumptions~\cite{poddar2025overview}, empirical path-loss formulas, or ray-tracing-based simulation. While these methods offer useful physical interpretations, they often require strong environmental priors~\cite{alkhateeb2019deepmimo,hoydis2024learning, gougeon2024ray, chen2026physically}, such as accurate geometry, material properties, or calibrated propagation parameters. These requirements are difficult to satisfy in practical indoor deployments, where floor plans may be incomplete, furniture layouts may change, and material properties are rarely known precisely. Moreover, analytical models often struggle to capture site-specific effects such as multipath, while ray tracing depends heavily on scene reconstruction quality and material calibration and can be computationally expensive for dense field reconstruction.

Recent learning-based sensing studies have also leveraged generative models to alleviate data scarcity, using RF data synthesis, missing-configuration generation, and cross-modal translation to improve sensing robustness~\cite{rfgenesis, sudokusens, xgait, 10.1145/3666025.3699351}. These methods highlight the value of generative learning for sensor data augmentation, but they mainly operate on task-specific sensing samples rather than field-level wireless propagation. In contrast, neural field methods aim to learn site-specific wireless representations directly from sparse measurements. Instead of explicitly tracing ray-scene interactions or relying on hand-crafted propagation models, these methods represent the wireless radiation field as a learnable continuous function optimized to reproduce measured wireless signals~\cite{nerf2, wrfgs, wrfgs-plus, newrf, 11152993, 11206262}. Inspired by neural radiance fields~\cite{mildenhall2021nerf, instant-ngp} and differentiable Gaussian scene representations~\cite{kerbl20233d}, neural wireless fields can flexibly encode propagation effects, including multipath, shadowing, scattering, and local attenuation. In particular, Gaussian-based wireless representations inherit the efficiency and spatial adaptivity of 3D Gaussian Splatting (3DGS), making them attractive for reconstructing wireless fields from sparse measurements. However, existing neural wireless models remain largely data-driven: their learned fields are constrained primarily by observed signal samples, while the physical laws governing wave propagation are only weakly reflected through architectural inductive biases. Under sparse measurement coverage, this can lead to interpolations that are numerically smooth but physically inconsistent in unobserved regions. For wireless propagation, such inconsistency is not merely a reconstruction artifact: a physically plausible field should follow wave-equation behavior in free space, attenuate or scatter near obstacles, and vary according to the propagation properties of the surrounding environment~\cite{ishimaru2017electromagnetic}.

A natural direction is to introduce physics-informed learning into neural radiation field reconstruction~\cite{raissi2017physics, cuomo2022scientific, lim2022maxwellnet}. However, directly appending a generic physics-informed neural network~(PINN)-style residual to an existing neural wireless model is insufficient. First, rendered wireless observations, such as received signal strength indicator (RSSI) or spatial spectra, are aggregated measurements rather than the underlying electromagnetic field governed by the wave equation; enforcing a differential equation directly on the rendered output therefore constrains the wrong physical quantity. Second, wave-equation constraints are valid only in appropriate propagation regions: free-space Helmholtz consistency should not be enforced uniformly inside walls, furniture, or other material obstacles, where propagation behavior differs from that of homogeneous air. Existing physics-regularized approaches often depend on externally provided geometry, such as computer-aided design (CAD) models, reconstructed meshes, or manually annotated floor plans~\cite{levie2021radiounet, 9771088}, which are expensive to obtain and frequently unavailable in practical deployments. An effective physics-informed wireless representation should therefore incorporate physical constraints in a scene-aware manner, rather than simply attaching a partial differential equation~(PDE) loss after rendering.

To address these limitations, we propose \textbf{PINN-GS}, a physics-informed 3DGS framework for wireless radiation field reconstruction from sparse measurements. The key distinction of PINN-GS is that physics supervision is integrated into the 3DGS pipeline, rather than applied as a standalone loss after rendering. PINN-GS attaches a lightweight physics branch to the shared Gaussian backbone, allowing the model to regularize the overall reconstructed wireless field while preserving the efficiency of Gaussian-based rendering. This branch predicts a wireless field over the scene and learns a soft free-space/obstacle mask directly from measurements. The wave field serves as the physical state variable on which Helmholtz consistency is imposed, while the mask implicitly captures obstacle-induced propagation properties by indicating where free-space propagation is valid and where material or blocked regions should relax the constraint. In this way, PINN-GS couples scene-aware physics regularization with the evolving Gaussian representation, without requiring CAD models, floor plans, or manually annotated environmental priors. Our contributions are summarized as follows: 

\begin{itemize}[leftmargin=*, topsep=0.25em, itemsep=0.25em, parsep=0pt]
    \item To the best of our knowledge, PINN-GS is the first wireless physics-informed 3DGS framework that integrates wave-equation regularization into the wireless 3D Gaussian Splatting pipeline.
    \item We design a scene-aware physics branch that predicts a wireless wave field and a soft medium mask for propagation-aware regularization. We further develop a training-time physics regularizer that improves reconstruction accuracy and physical consistency. 
    \item Experiments on real-world wireless sensing benchmarks show that PINN-GS consistently outperforms existing neural wireless field reconstruction methods across RSSI and RFID spectrum reconstruction tasks. Our code is available at \url{https://github.com/NorahCadeJS/PINN-GS}.
\end{itemize}

\section{Preliminaries}

\subsection{Physics-Informed Neural Networks}
PINNs incorporate governing physical laws into neural network training through equation-based regularization~\cite{raissi2017physics,cuomo2022scientific}. Instead of learning solely from labeled observations, a PINN constrains the learned representation to satisfy known differential equations, thereby encouraging physically consistent predictions in regions where measurements are sparse or noisy.

Let $u_{\theta}(\mathbf{x})$ denote a neural approximation of a physical field defined over spatial coordinates $\mathbf{x}$. Given a governing differential operator $\mathcal{N}[\cdot]$, the physical residual is
$\mathcal{R}(\mathbf{x}) = \mathcal{N}[u_{\theta}](\mathbf{x}).$
The physics-informed loss is then evaluated over a set of collocation points $\{\mathbf{x}_{i}\}_{i=1}^{N}$:
\begin{equation}
\mathcal{L}_{\mathrm{phys}} = \frac{1}{N} \sum_{i=1}^{N} \left| \mathcal{R}(\mathbf{x}_{i}) \right|^{2}.
\end{equation}
This term is combined with a data-fitting objective to form the overall loss
\begin{equation}
\mathcal{L} = \mathcal{L}_{\mathrm{data}} + \lambda_{\mathrm{phys}} \mathcal{L}_{\mathrm{phys}},
\end{equation}
where $\lambda_{\mathrm{phys}}$ balances measurement supervision against physical regularization. By penalizing violations of the governing equation, PINNs provide a principled mechanism for suppressing physically implausible solutions and improving generalization beyond observed samples.

\subsection{Physics-Based Wave Prior}
Wireless propagation is governed by electromagnetic wave physics. For a wireless signal at carrier frequency $f$, the time-harmonic field describes the spatial variation of the transmitted wave as it propagates through the environment. This formulation captures key propagation effects such as attenuation, phase variation, and interference, and provides a natural basis for imposing wave-equation consistency on the reconstructed wireless field. In a homogeneous medium, the time-harmonic field satisfies the Helmholtz equation~\cite{ishimaru2017electromagnetic}:
\begin{equation}
\nabla^{2}u(\mathbf{x}) + k^{2}u(\mathbf{x}) = 0,
\end{equation}
where $\nabla^{2}$ is the spatial Laplacian, $k = 2\pi/\lambda$ is the wavenumber, and $\lambda$ is the wavelength. This equation characterizes the spatial evolution of a wave at a fixed frequency and can be regarded as the frequency-domain form of the wave equation under the time-harmonic assumption.

The Helmholtz equation provides a natural physical prior for wireless field reconstruction, as it constrains how the electromagnetic field varies in free-space propagation regions. Enforcing consistency with this equation encourages the reconstructed field to follow physically plausible wave behavior rather than relying purely on interpolation between measured samples.

\section{Framework}

\begin{figure}[htbp]
\centering
\includegraphics[width=\linewidth]{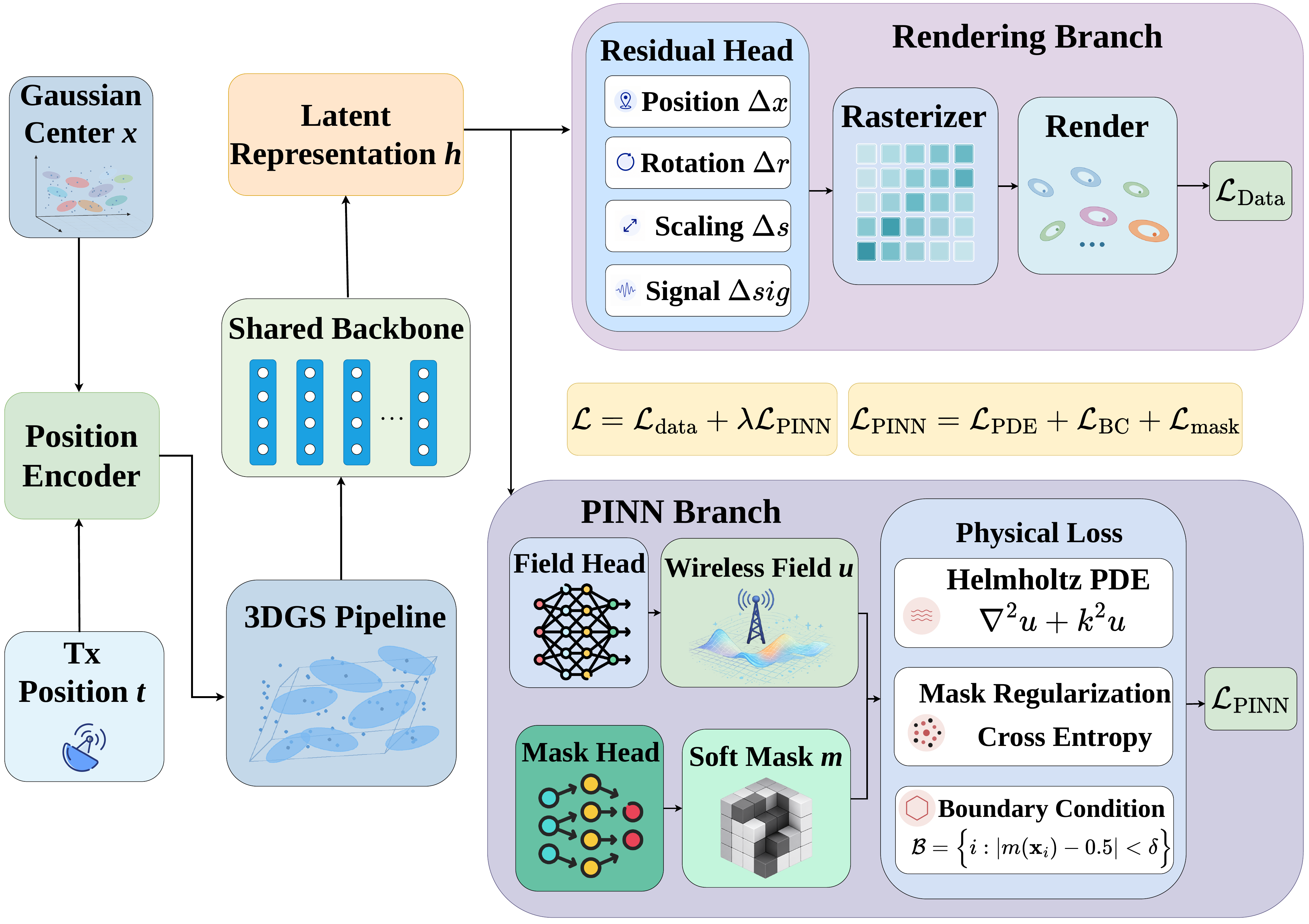}
\caption{Model overview.}
\label{figure_model}
\vspace{-3pt}
\end{figure}

\subsection{Overview}
Figure~\ref{figure_model} presents an overview of PINN-GS. We represent the propagation environment with a set of $M$ anisotropic 3D Gaussians $\mathcal{G}_i = (\mathbf{x}_i, \mathbf{r}_i, \mathbf{s}_i, \alpha_i, \mathbf{sig}_i)$, where $\mathbf{x}_i$, $\mathbf{r}_i$, $\mathbf{s}_i$, $\alpha_i$, and $\mathbf{sig}_i$ denote the center, rotation, scale, opacity, and signal attribute of the $i$-th primitive. Given a transmitter position $\mathbf{t} \in \mathbb{R}^3$, each pair $(\mathbf{x}_i, \mathbf{t})$ is encoded and mapped by a shared deformation backbone $f_\theta$ to a transmitter-aware latent feature $\mathbf{h}_i$, which serves as the shared representation for both branches. From this shared latent space, PINN-GS instantiates two forward-decoupled branches, as illustrated in Fig.~\ref{figure_model}. The rendering branch predicts per-Gaussian residuals that deform the primitives and are rasterized into a wireless observation, supervised by the data-fitting loss $\mathcal{L}_{\mathrm{data}}$. In parallel, the physics-informed branch predicts a wave field and a soft free-space/obstacle mask, from which the physics objective $\mathcal{L}_{\mathrm{PINN}}$ imposes wave-equation consistency over the scene. Crucially, the two branches share parameters only through the backbone and never exchange forward signals: the rasterizer does not consume the physical field, and the wave operator does not consume the rendered output. Their coupling arises solely through back-propagation into $\mathbf{h}_i$, so that the learned representation is shaped jointly by data fidelity and propagation physics.

\subsection{3DGS Pipeline and Shared Backbone}
As shown in Fig.~\ref{figure_model}, our 3DGS pipeline takes the Gaussian center $\mathbf{x}_i$ and the transmitter position $\mathbf{t}$ as inputs and renders them into a wireless observation through a shared backbone and a rendering branch. To condition the scene representation on the transmitter location, we first map $\mathbf{x}_i$ and $\mathbf{t}$ through a positional encoder, denoted $\gamma(\mathbf{x}_i)$ and $\gamma(\mathbf{t})$. The two embeddings are concatenated and passed to a shared backbone that produces a transmitter-aware latent feature $\mathbf{h}_i$ for each Gaussian primitive. This design follows the general philosophy of the WRF-GS family~\cite{wrfgs,wrfgs-plus}, where the wireless scene is represented by a set of learnable Gaussian primitives whose attributes adapt to different transmitter configurations. Indoor propagation depends on both the spatial distribution of scene structures and their effects on signal attenuation. The Gaussian representation provides a compact way to capture these factors. The positions and scales of the primitives describe where scene structures are represented, while their learned opacities provide a latent proxy for blocking and attenuation effects.

Unlike prior models that use separate feature extractors for different objectives, we employ a single shared backbone to provide a common latent representation $\mathbf{h}$ for both the rendering branch and the physics-informed branch. As illustrated in Fig.~\ref{figure_model}, $\mathbf{h}$ is the sole point of contact between the two branches: it is consumed by the residual heads that drive differentiable splatting, and by the field and mask heads that impose physics regularization. Sharing the backbone therefore couples data-driven wireless rendering with physics supervision, encouraging the rendered observations and the learned physical field to be derived from a consistent scene representation rather than from two independent latent spaces.

\subsection{Rendering Branch}
The rendering branch maps the shared latent feature $\mathbf{h}_i$ to the measurable wireless observation through differentiable Gaussian splatting. In Fig.~\ref{figure_model}, $\mathbf{h}_i$ is fed to a set of residual heads that predict per-Gaussian updates $(\Delta\mathbf{x}_i, \Delta\mathbf{r}_i, \Delta\mathbf{s}_i, \Delta\mathbf{sig}_i)$ to the primitive's center, rotation, scale, and signal attribute. These Gaussian residuals deform the primitive geometry and adjust its response before rasterization, enabling each Gaussian to represent not only static scene structure but also transmitter-dependent propagation effects such as blockage, reflection, and scattering. This conditional deformation is important for wireless modeling, since the effective contribution of a scatterer or obstacle can vary significantly with transmitter location even when the physical environment is unchanged.

The updated Gaussian primitives are then projected and alpha-composited by a differentiable rasterizer to produce the wireless observation $\hat{\mathbf{y}}(\mathbf{t})$ rendered by the Gaussian splatting branch. This rendering process is supervised by the data-fitting loss $\mathcal{L}_{\mathrm{data}}$, which measures the discrepancy between the rendered wireless observations and the corresponding measured signals. Specifically, the rendering branch is optimized with an L2 reconstruction objective over all measurements:
$\mathcal{L}_{\mathrm{data}} =
\frac{1}{N}
\sum_{j=1}^{N}
\left|
\hat{\mathbf{y}}(\mathbf{t}_j)-\mathbf{y}_j
\right|_2^2 ,$
where $N$ denotes the total number of wireless measurements, $\mathbf{t}_j$ represents the measurement position or query coordinate, and $\mathbf{y}_j$ is the corresponding measured wireless target. In this way, the rendering branch serves as the primary observation-generation module of the framework, transforming the Gaussian representation into measurable wireless quantities while preserving end-to-end differentiability for joint optimization with the physics-informed branch.

\subsection{PINN Branch}

The physics-informed branch attaches two lightweight heads to the shared backbone: a field head that predicts a local wave field, and a mask head that infers a soft free-space/obstacle partition. These heads are used only during training and are not part of the rendering path at inference time. Their purpose is to impose propagation-aware constraints on the shared latent representation, so that the learned Gaussian scene is not only consistent with measured wireless observations, but also compatible with the governing wave equation in physically valid regions.

Unlike the rendering branch, which maps the Gaussian scene representation to observable wireless quantities, the physics-informed branch operates on the latent field defined over the scene. This design allows the model to regularize regions beyond sparsely sampled receiver locations, providing additional physical guidance for wireless field reconstruction without requiring explicit geometry, material labels, or dense field measurements.

\paragraph{Wireless field.}

We model the time-harmonic electromagnetic field induced by the transmitter as a function $u(\mathbf{x_i},\mathbf{t})$. At each Gaussian center, the field is predicted by the field head $\phi_{\mathrm{field}}$ from the shared latent feature:
\begin{equation}
u(\mathbf{x_i},\mathbf{t})=\phi_{\mathrm{field}}\!\left(\mathbf{h}(\mathbf{x_i},\mathbf{t})\oplus\gamma(\mathbf{x_i})\right).
\end{equation}
For brevity, we omit the fixed transmitter argument \(t\) in the following physics losses. The positional skip connection $\gamma(\mathbf{x})$ preserves fine-grained spatial information that may be smoothed by the shared backbone. This is important for wireless propagation, where the field can exhibit spatial variation due to phase rotation, path-length differences, and multipath interference. The representation naturally captures both amplitude and phase: the amplitude reflects attenuation and signal strength variation, while the phase encodes propagation delay and constructive or destructive interference among paths. The field head represents the complex wave field with two output channels corresponding to its real and imaginary components. These channels are used only by the physics-informed branch and are not passed to the rendering branch. This design allows the Helmholtz equation to constrain the latent wave field while keeping the physics and rendering paths forward-decoupled.

The predicted field is not directly consumed by the rasterizer and is not supervised as an additional observation target. Instead, it provides the physical state on which the wave equation is evaluated. Because the field head and the rendering branch share the same backbone, gradients from the physical loss regularize the latent representation used for rendering. The learned representation is therefore guided by both measurement fidelity and wave-consistent spatial structure, especially in regions with sparse observations.

\begin{figure}[htbp]
\centering
\includegraphics[width=\linewidth]{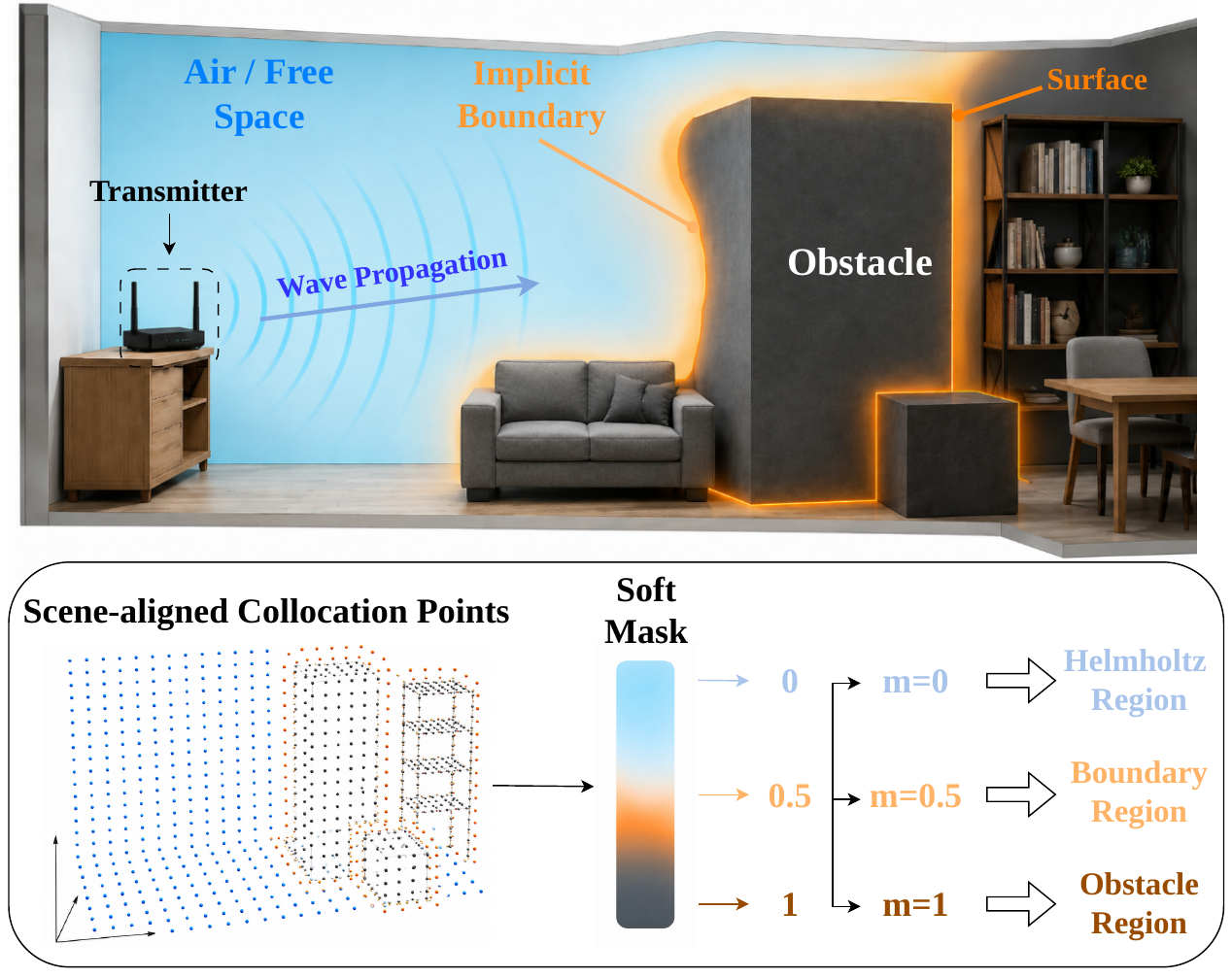}
\caption{Soft mask for implicit propagation geometry.}
\vspace{-3pt}
\label{figure_mask}
\end{figure}

\paragraph{Helmholtz PDE}

In homogeneous free space away from sources, a time-harmonic electromagnetic field satisfies $\nabla^2 u(\mathbf{x}) + k^2 u(\mathbf{x}) = 0$. The transmitter lies inside the region we constrain, so the governing equation there is the inhomogeneous form $\nabla^2 u(\mathbf{x}) + k^2 u(\mathbf{x}) = -q(\mathbf{x})$, where $q$ is a smoothed point source centred at the transmitter. In indoor wireless environments, however, this free-space assumption is valid only in air regions and should not be enforced inside walls, furniture, or other material obstacles. We use the learned mask $m(\mathbf{x})$ to derive an air probability that weights the PDE residual: $ w_{\mathrm{air}}(\mathbf{x})=\max\!\left(1-m(\mathbf{x}),\epsilon_{\mathrm{air}}\right),$
where $\epsilon_{\mathrm{air}}$ prevents the gradient from vanishing completely when the mask saturates. The resulting PDE loss is
\begin{equation}
\mathcal{L}_{\mathrm{PDE}} =
\frac{\sum_{i\in\mathcal{C}} w_{\mathrm{air}}(\mathbf{x}_i)\left|\nabla^2 u(\mathbf{x}_i) + k^2 u(\mathbf{x}_i) + q(\mathbf{x}_i)\right|^2}
     {\sum_{i\in\mathcal{C}} w_{\mathrm{air}}(\mathbf{x}_i)\left\|q(\mathbf{x_i})\right\|^2},
\end{equation}
where $\mathcal{C}$ is a collocation set of Gaussian centers by opacity-weighted sampling at each step. This loss encourages the predicted field to follow free-space wave propagation in regions where the mask indicates air, while allowing material regions to deviate from the homogeneous-medium equation. Evaluating the residual at Gaussian centers provides physics supervision beyond the measured receiver locations.

We normalize the PDE residual by the source strength to obtain a source-independent reference for detecting the trivial solution. When $u\equiv0$, the field-dependent terms vanish and the normalized residual evaluates to exactly 1, regardless of the source amplitude. Therefore, a residual below 1 indicates that the learned field does not collapse to the zero-field solution and better satisfies the source-driven wave equation.

Several design choices reduce degenerate solutions. First, the source term makes $u\equiv0$ inconsistent with the governing equation. Second, the mask regularization prevents the model from suppressing the PDE residual by assigning the entire scene to the obstacle region. Third, the wavenumber is scaled to the scene so that the corresponding oscillation remains resolvable at the available collocation density.

We compute the Laplacian with automatic differentiation rather than finite differences, so the PDE residual does not require a regular spatial grid. Collocation points are sampled from Gaussian centers according to their opacity.

\paragraph{Soft mask.}
The second head $\phi_{\mathrm{mask}}$ predicts a continuous soft medium mask
\begin{equation}
m(\mathbf{x},\mathbf{t}) = \sigma\!\left(\phi_{\mathrm{mask}}(\mathbf{h}(\mathbf{x},\mathbf{t}))\right) \in [0,1],
\end{equation}
where $\sigma(\cdot)$ is the activation function that maps the mask head output to the unit interval. This mask acts as a learned implicit medium map over the scene, as illustrated in Fig.~\ref{figure_mask}. Rather than predicting a single occupancy value, the mask softly partitions space into three physically meaningful regimes: $m\!\to\!0$ corresponds to the blue air volume, where propagation follows the free-space Helmholtz equation; $m\!\to\!1$ corresponds to the dark obstacle volume, where blocking, absorption, and material-dependent attenuation invalidate the homogeneous-medium assumption; and the intermediate band $m\!\approx\!0.5$, shown as the orange interface, marks the implicit air--material boundary along which the field is expected to decay. Crucially, this decomposition is inferred jointly from the wireless measurements and the physics regularization, as annotated in Fig.~\ref{figure_mask}, so that PINN-GS recovers the propagation geometry without CAD models, meshes, floor plans, or manual material labels.

The mask is a structural component of the physics branch rather than an auxiliary prediction target, because it determines where each physical constraint applies. The Helmholtz equation holds only in homogeneous media, yet indoor environments contain walls and furniture that block or scatter the field. Enforcing a free-space residual uniformly would thus create a self-contradictory objective: the rendering objective favors attenuated responses in obstacle regions, while a uniformly applied free-space physics loss would encourage free-space behavior in the same regions. The mask resolves this conflict by gating the constraints in agreement with the three regimes of Fig.~\ref{figure_mask}, applying the Helmholtz residual in air, suppressing it inside obstacles, and activating a boundary attenuation regularization only along the learned interface.

We keep the mask soft for both optimization and modeling reasons. A hard binary mask would make the partition non-differentiable and block gradients almost everywhere, preventing the inferred geometry from adapting during training, and would render the boundary band around $m\!=\!0.5$ ill-posed. The sigmoid mask instead preserves differentiability, letting the data-fitting loss, Helmholtz residual, and boundary regularization jointly update both the wave field and the implicit medium structure. It therefore serves as a differentiable surrogate for the underlying propagation geometry, enabling physics-informed learning in indoor environments without explicit scene supervision.

\begin{figure*}[t]
  \centering
  \setlength{\tabcolsep}{0pt}
  \renewcommand{\arraystretch}{0.95}

  \setlength{\colw}{0.103\textwidth}
  \setlength{\colgap}{0.006\textwidth}

  \makebox[\textwidth][c]{%
  \begin{tabular}{@{}
    c@{\hspace{\colgap}}
    c@{\hspace{\colgap}}
    c@{\hspace{\colgap}}
    c@{\hspace{\colgap}}
    c@{\hspace{\colgap}}
    c@{\hspace{\colgap}}
    c@{\hspace{\colgap}}
    c@{\hspace{\colgap}}
    c@{}}

    \colstack{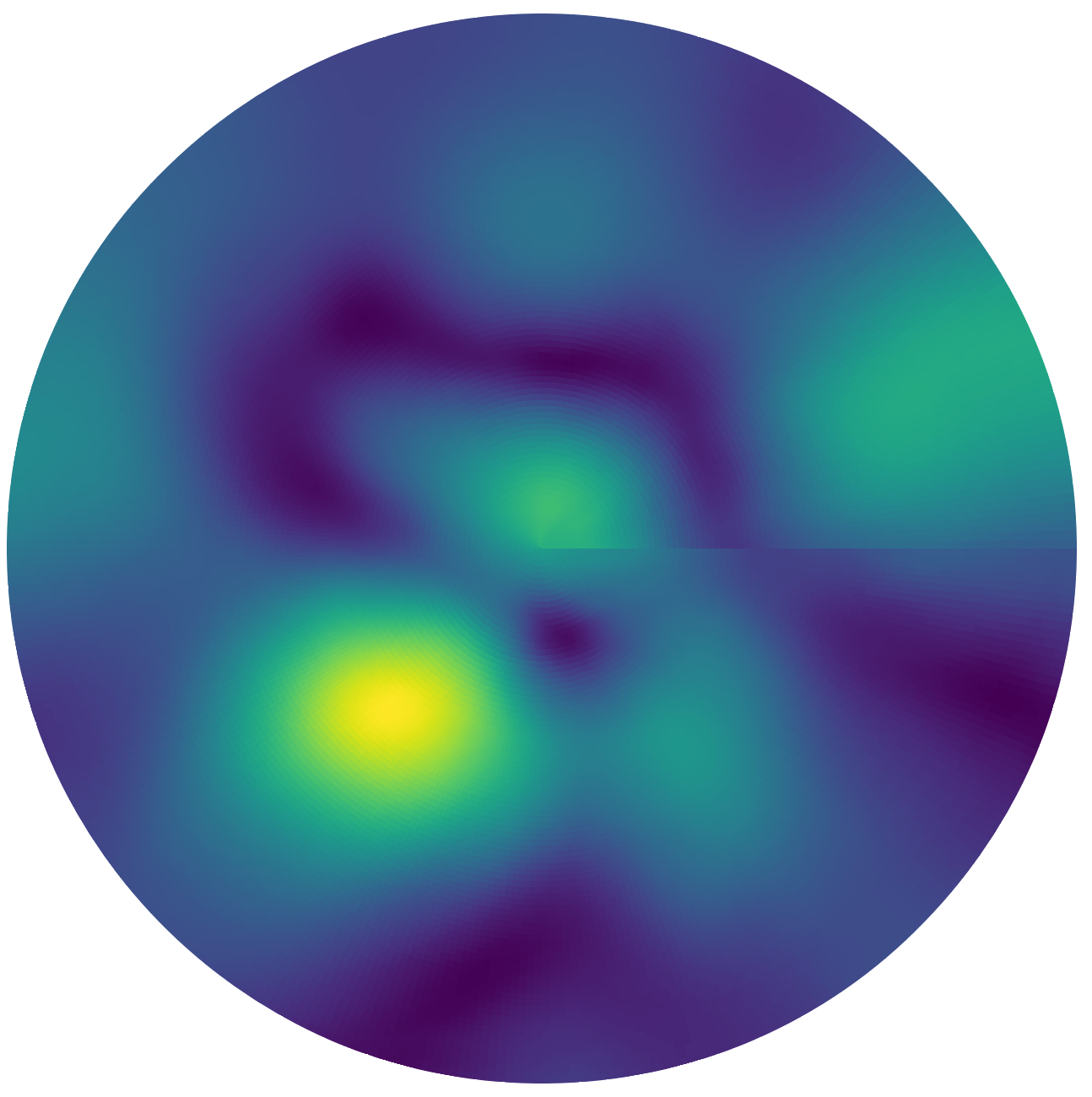}{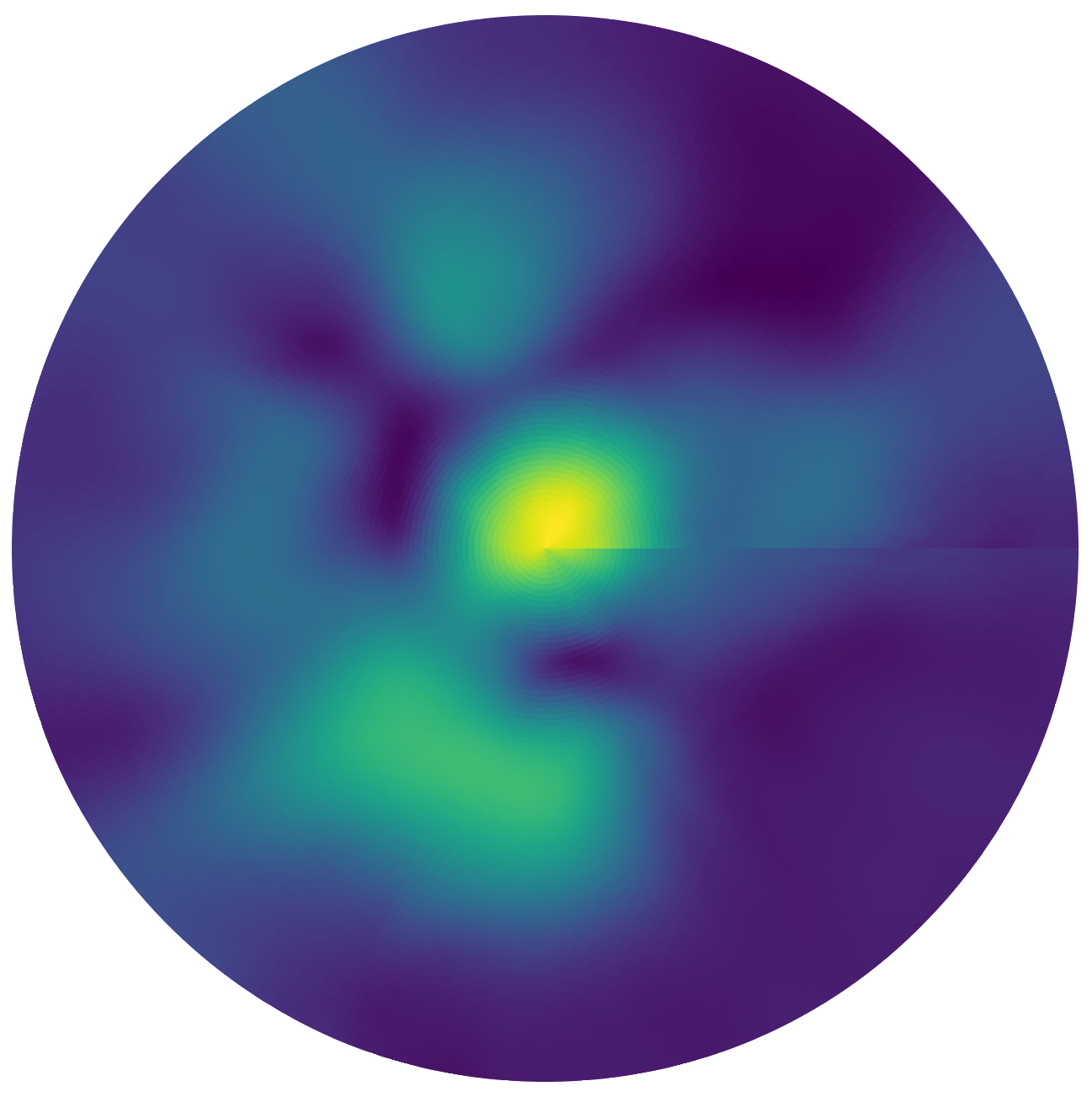}{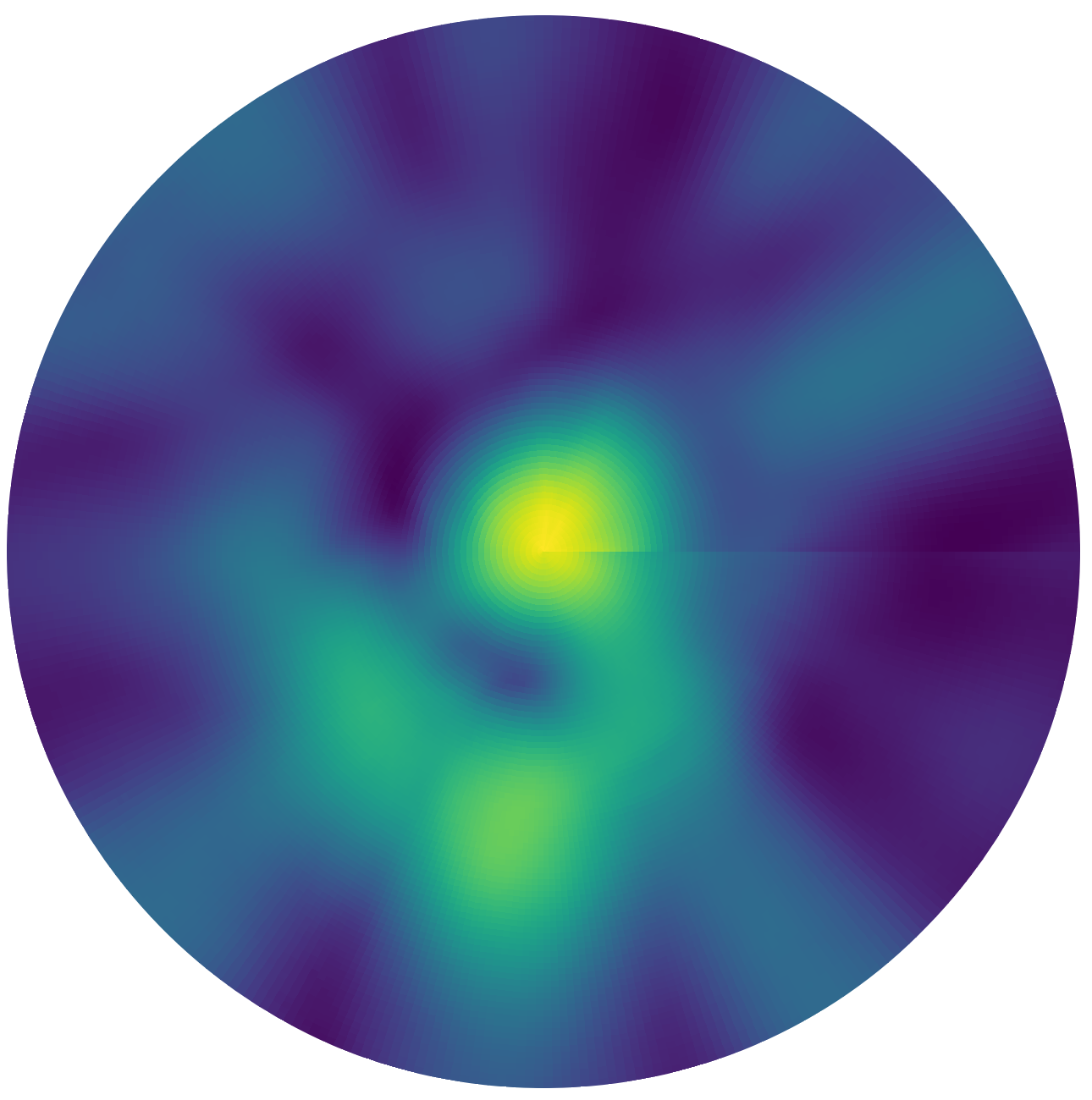} &
    \colstack{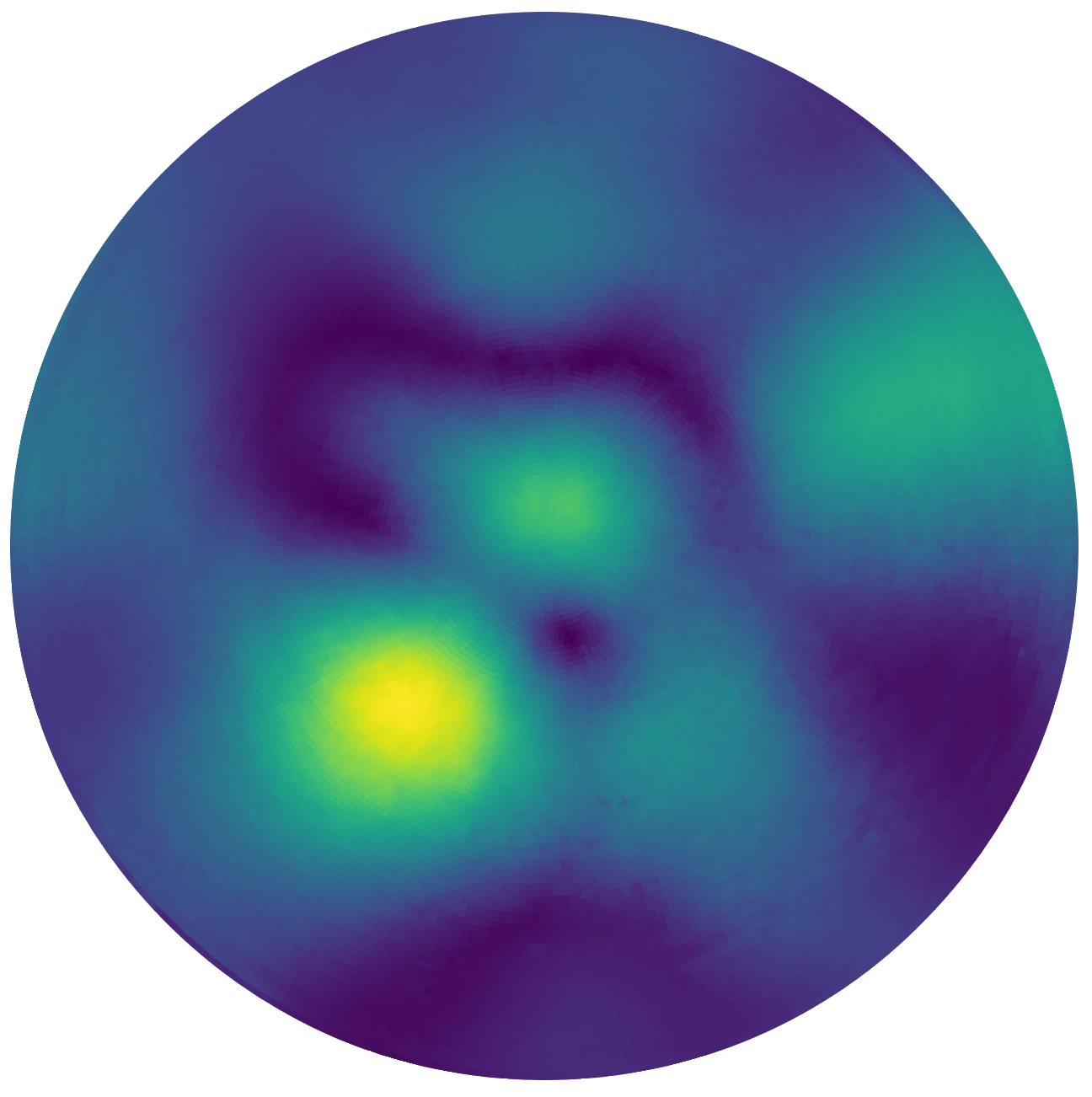}{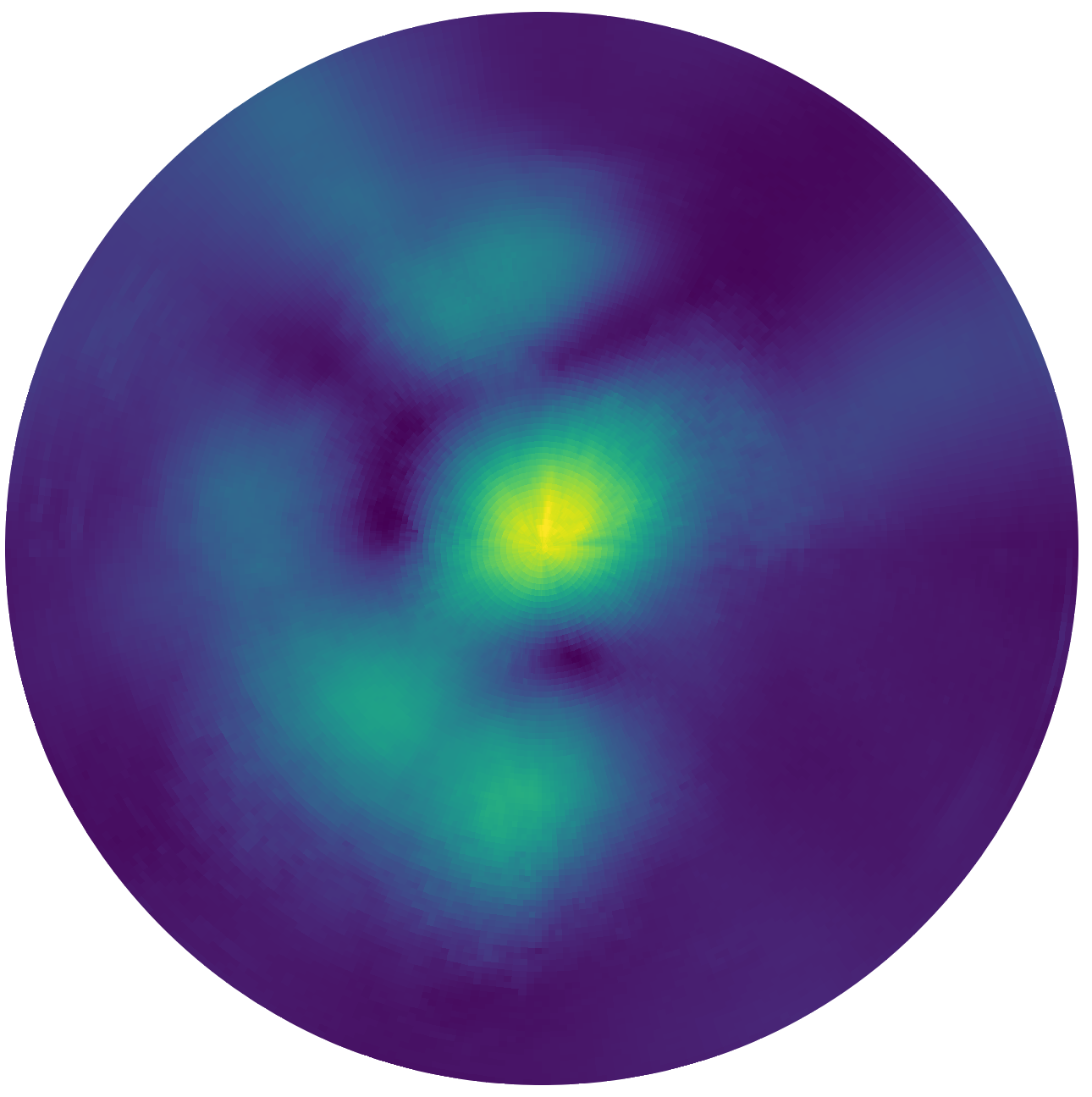}{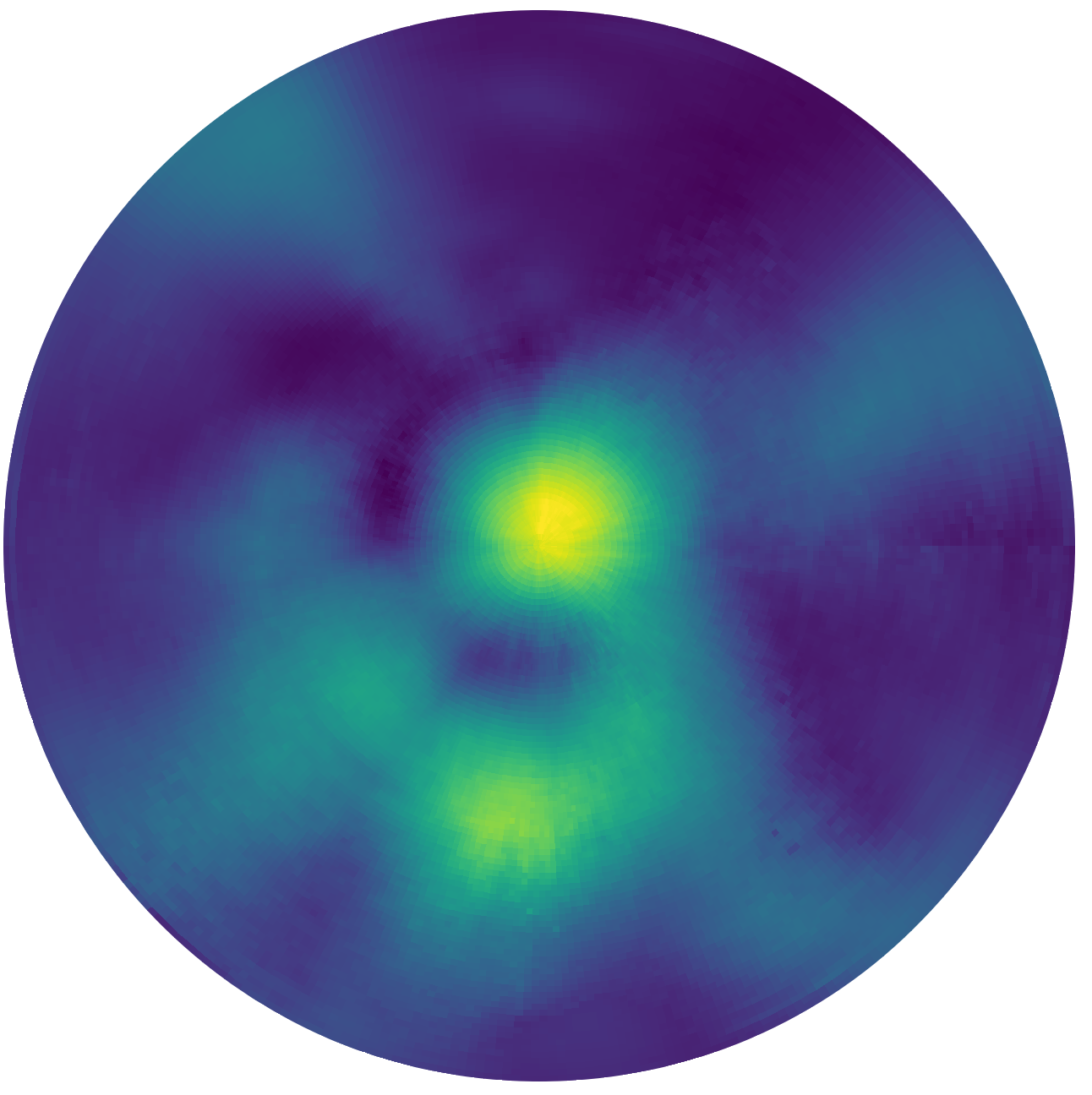} &
    \colstack{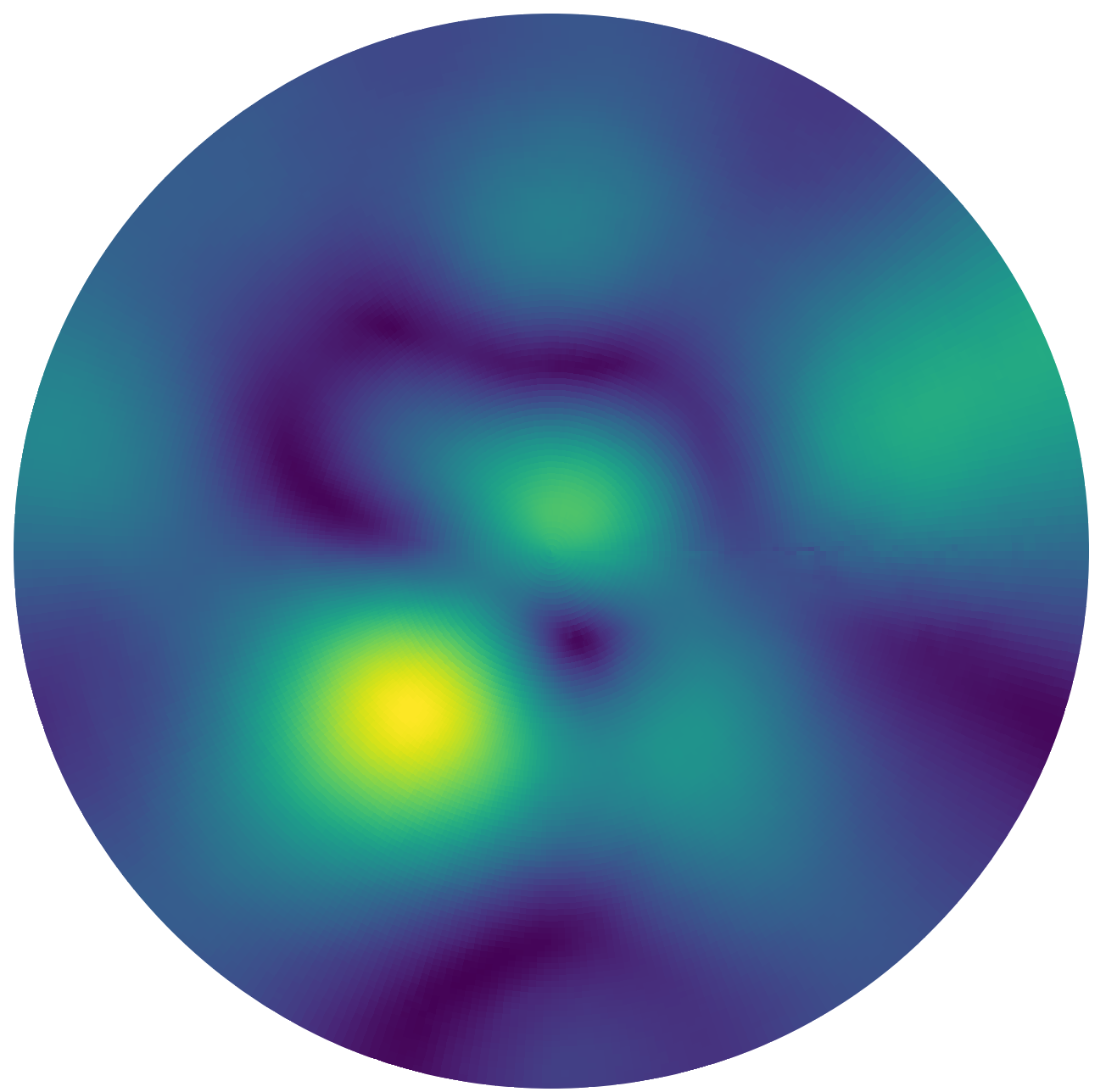}{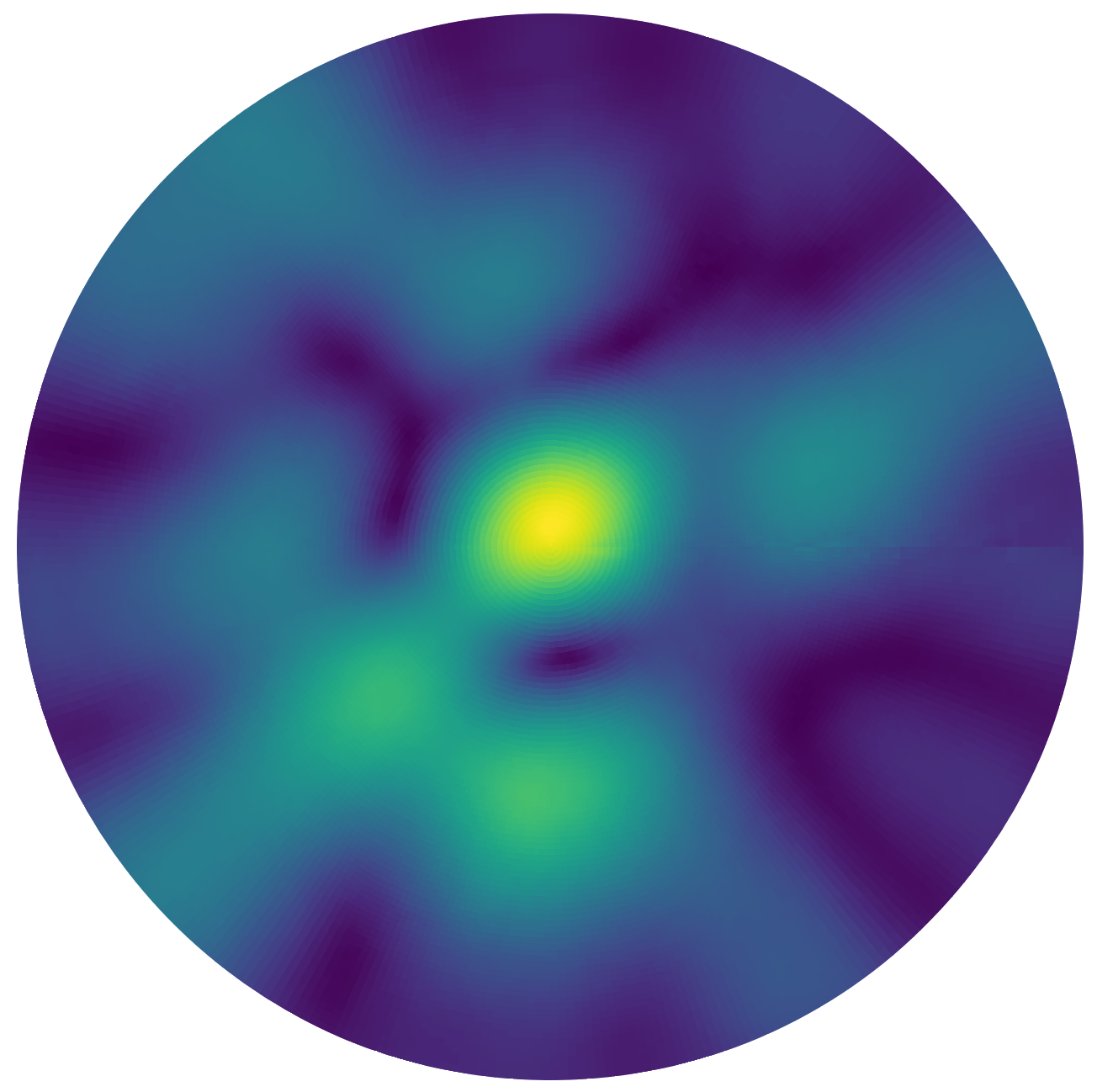}{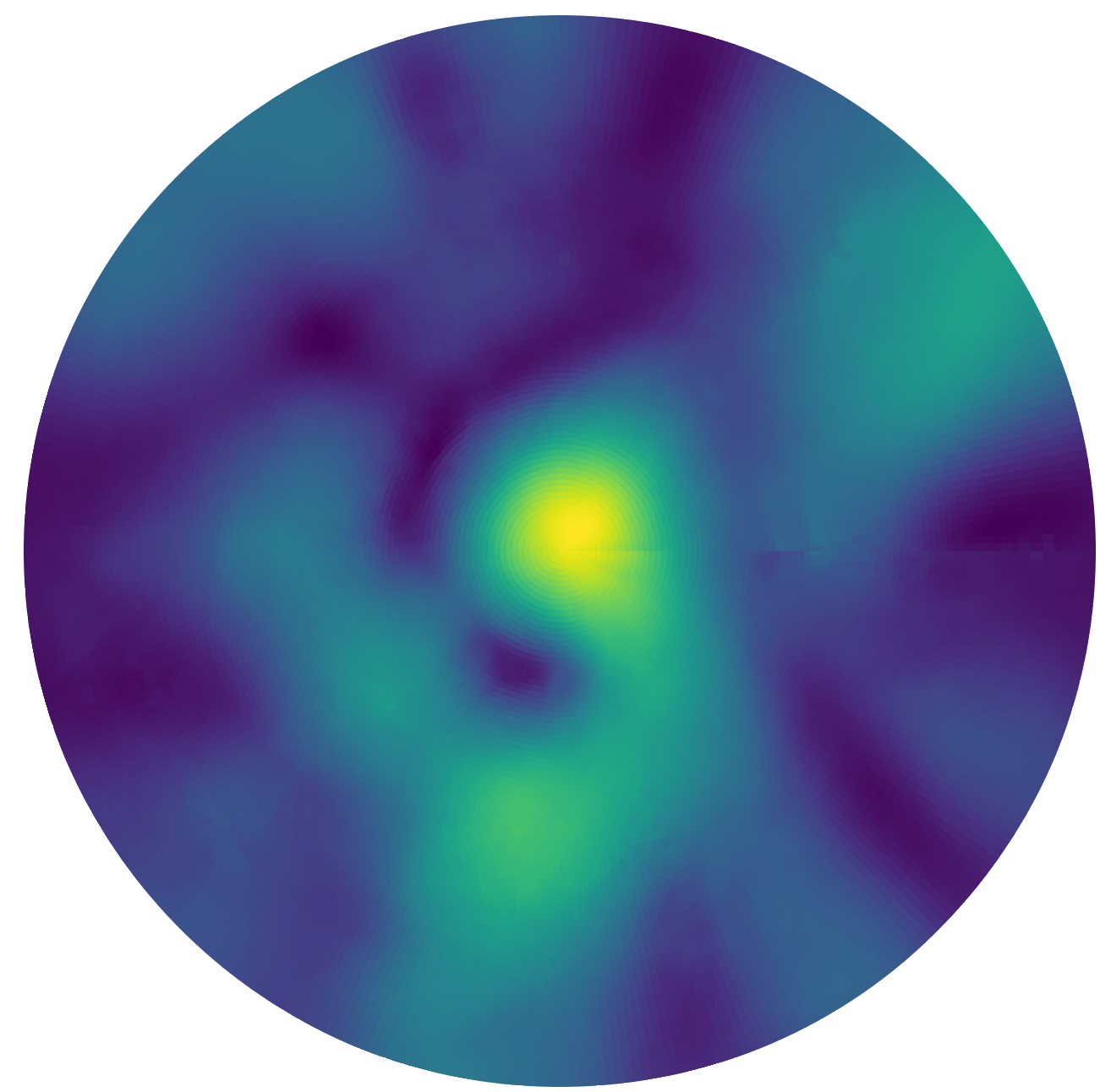} &
    \colstack{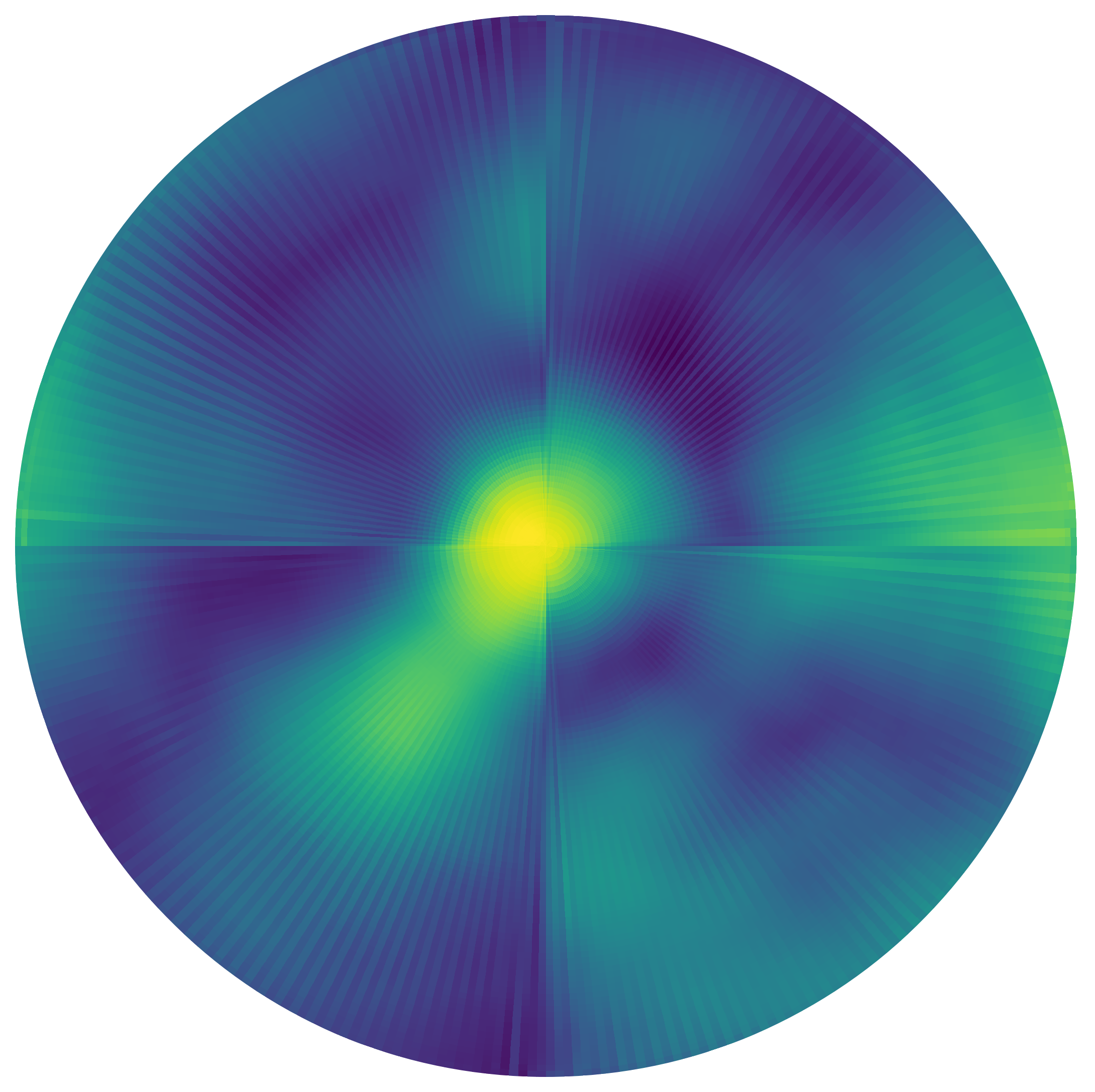}{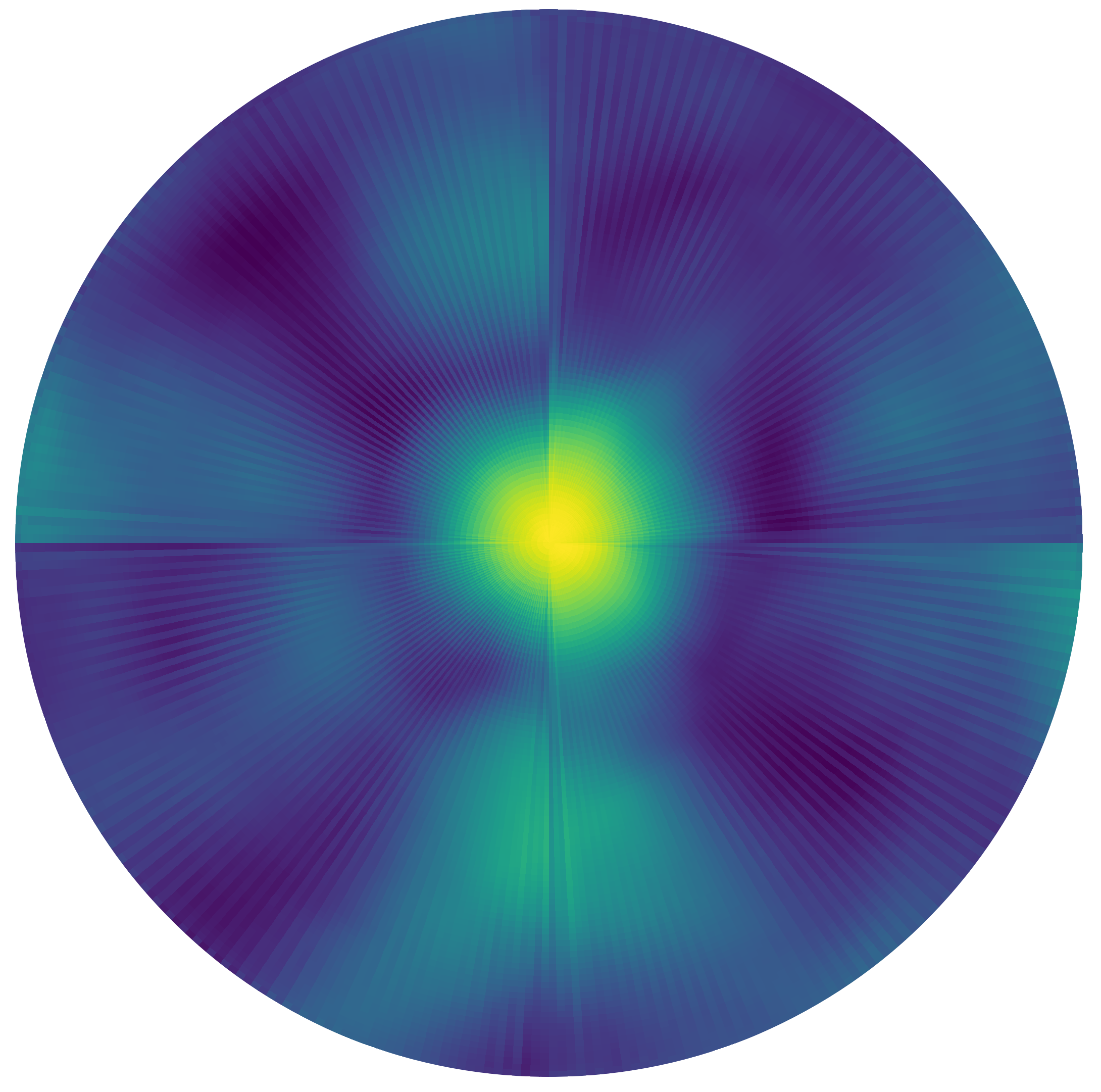}{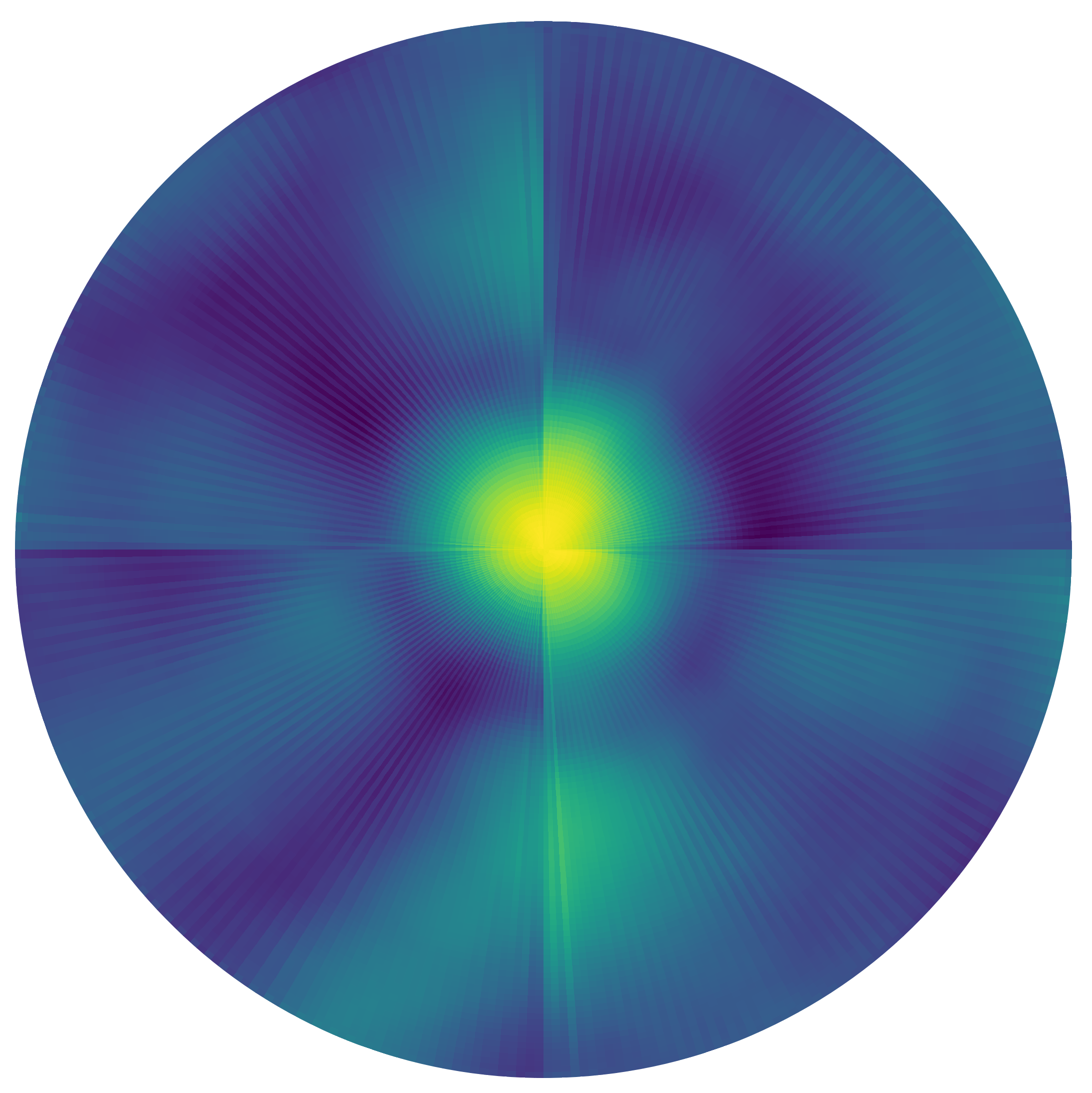} &
    \colstack{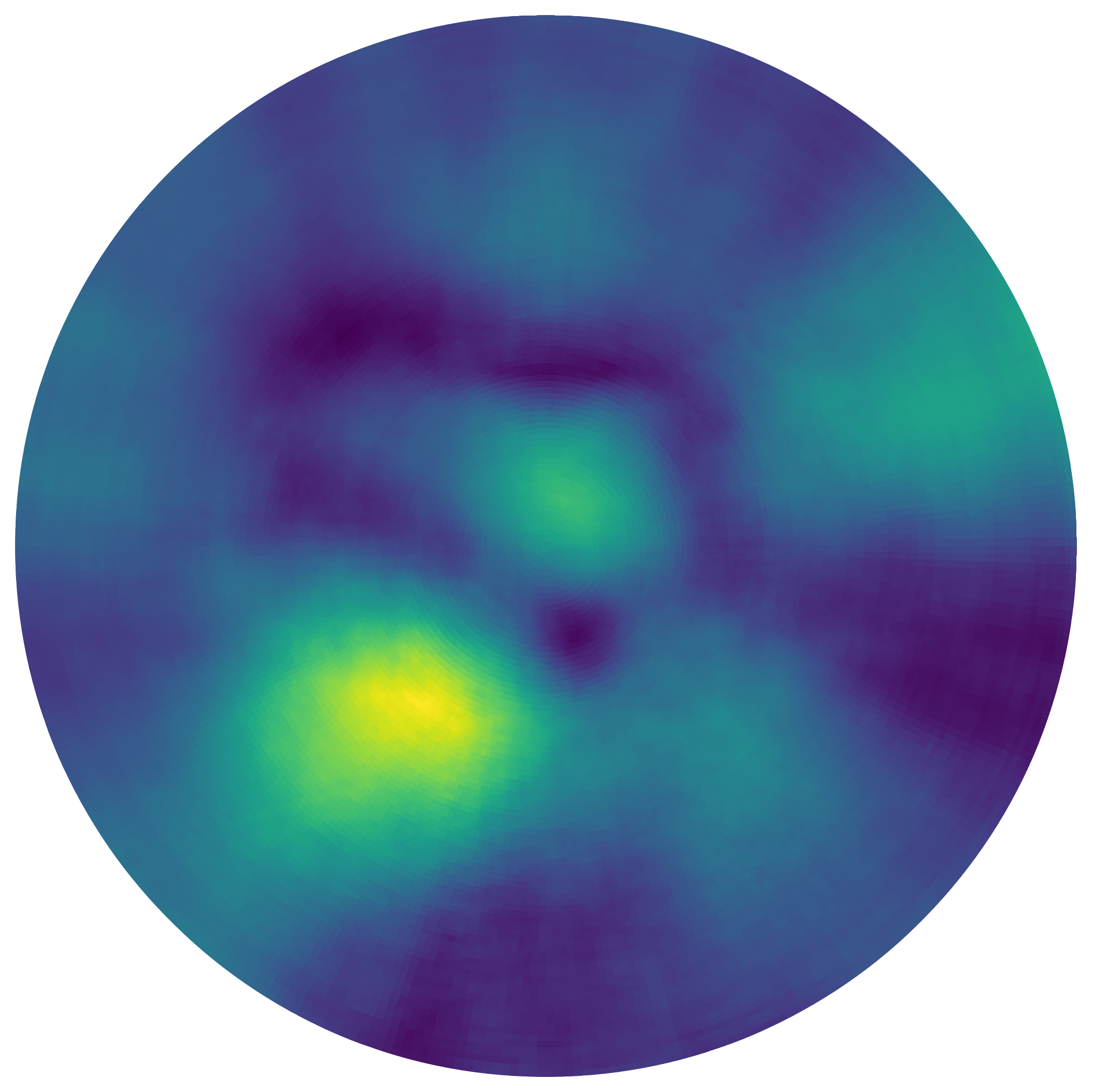}{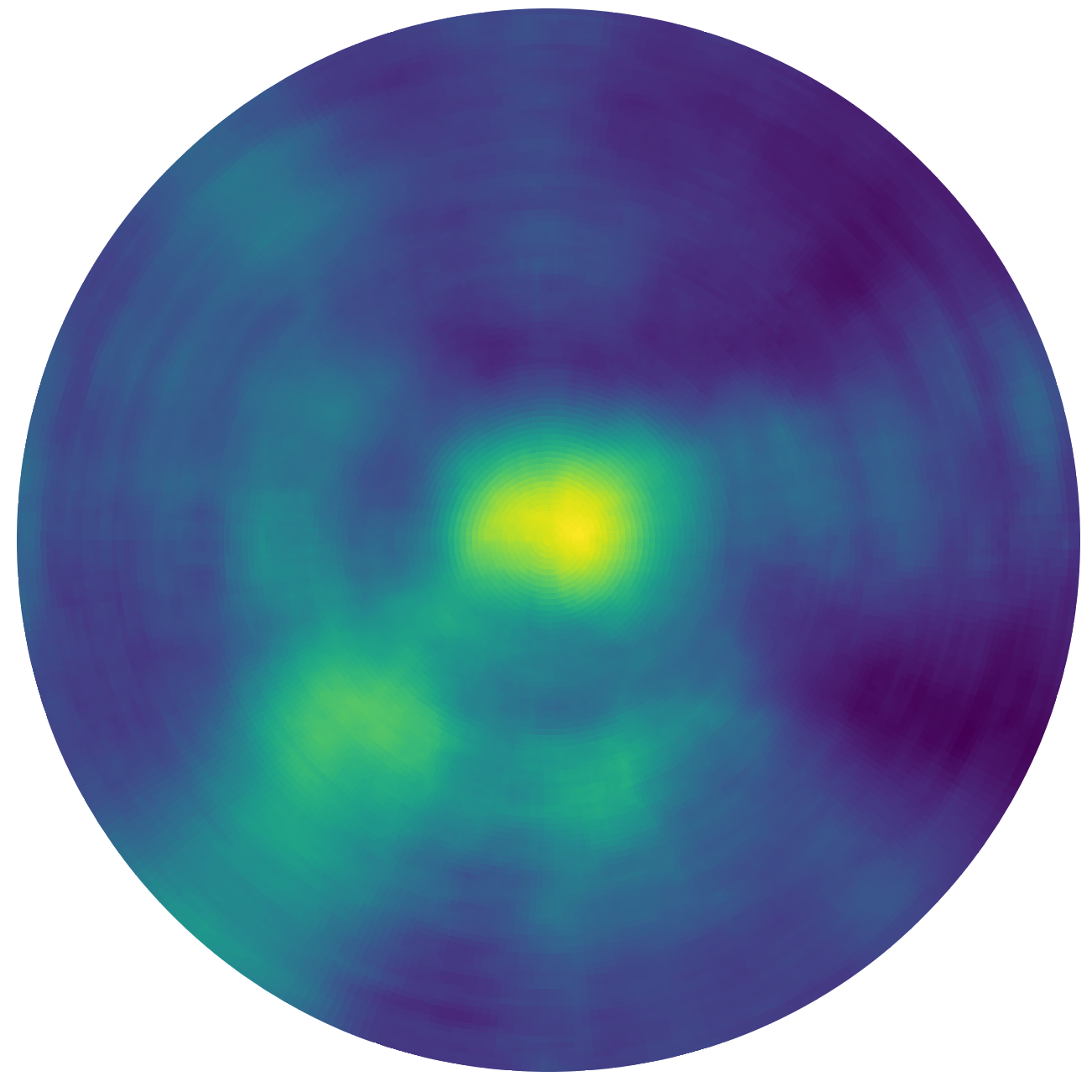}{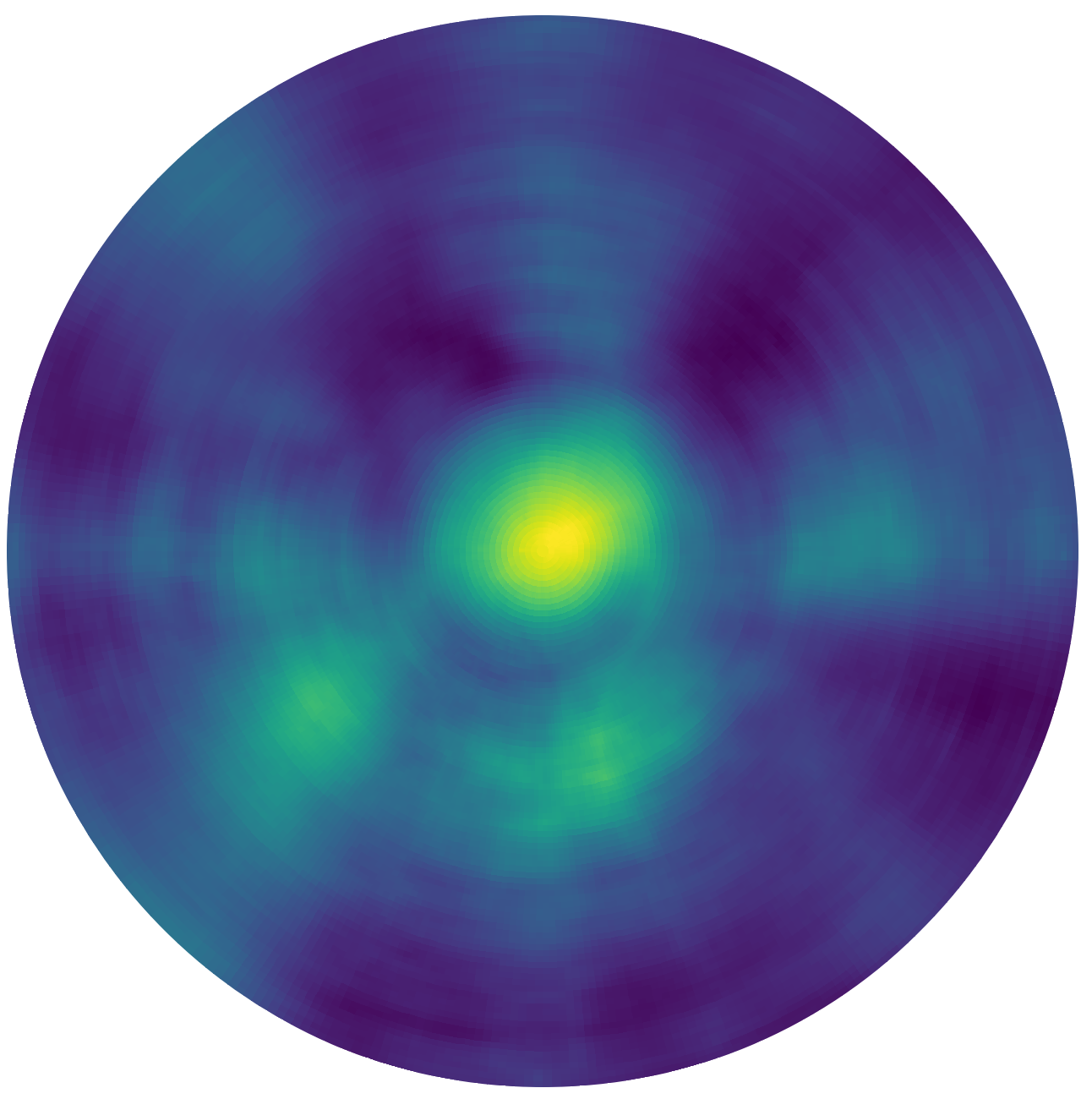} &
    \colstack{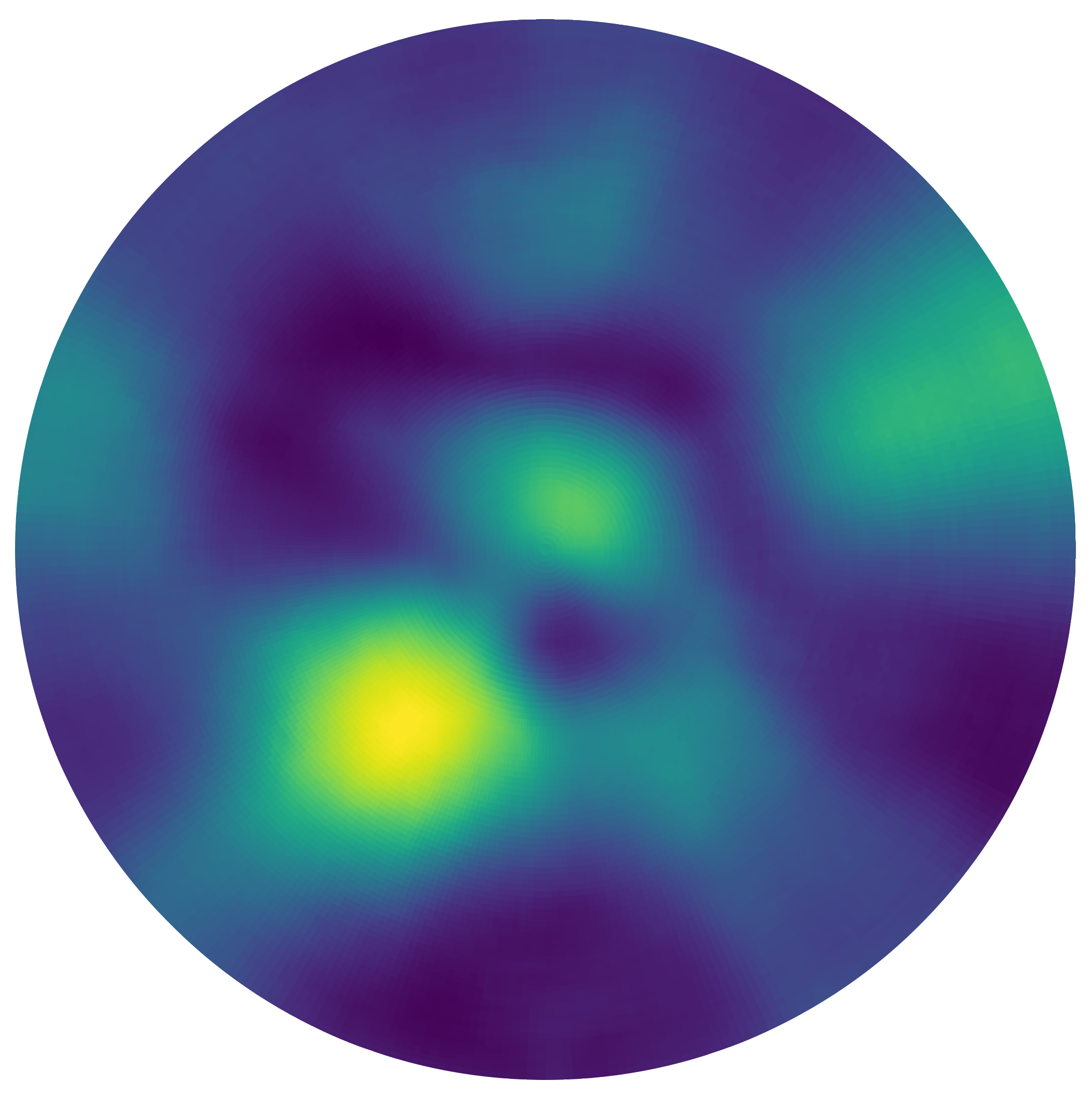}{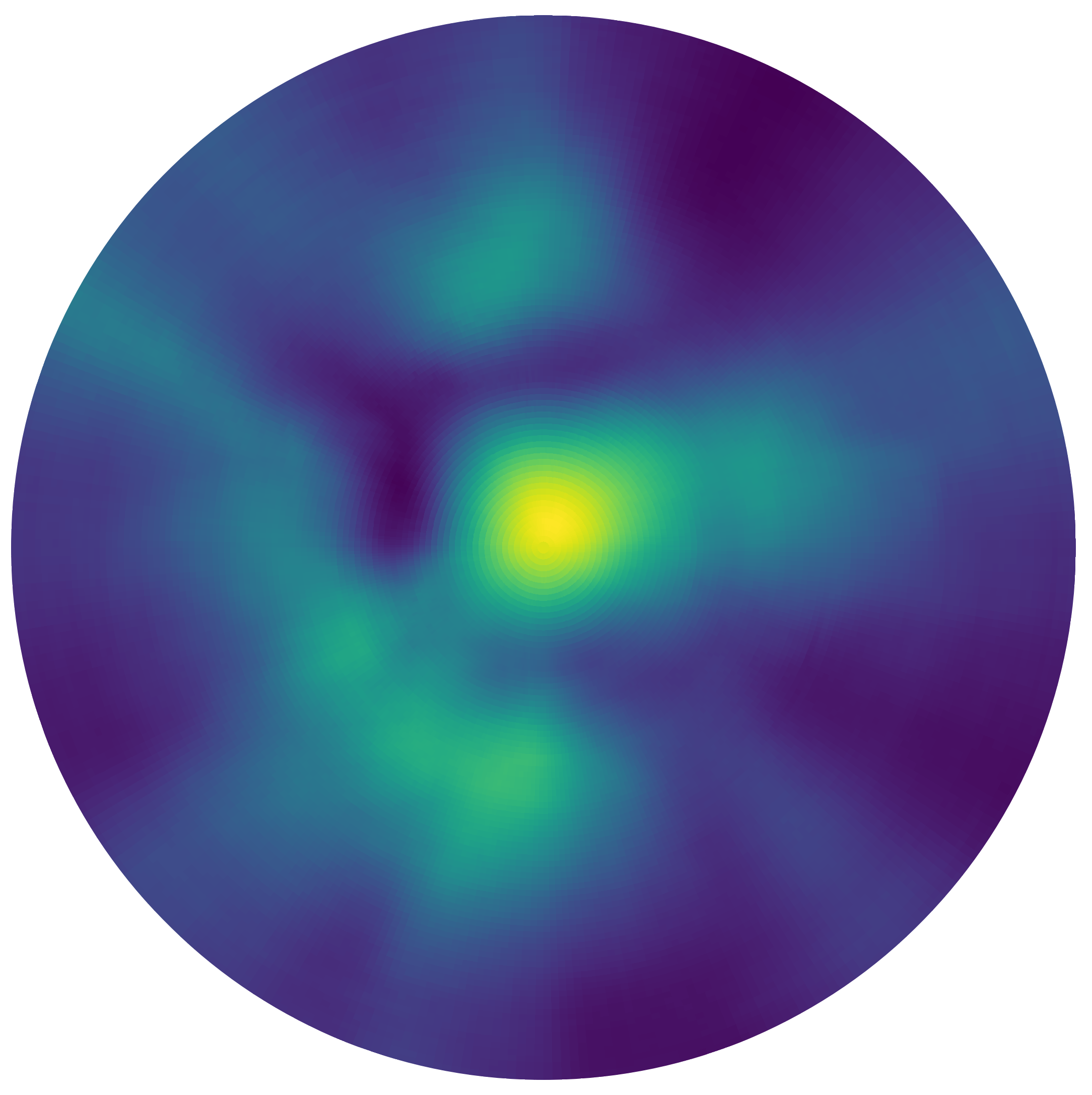}{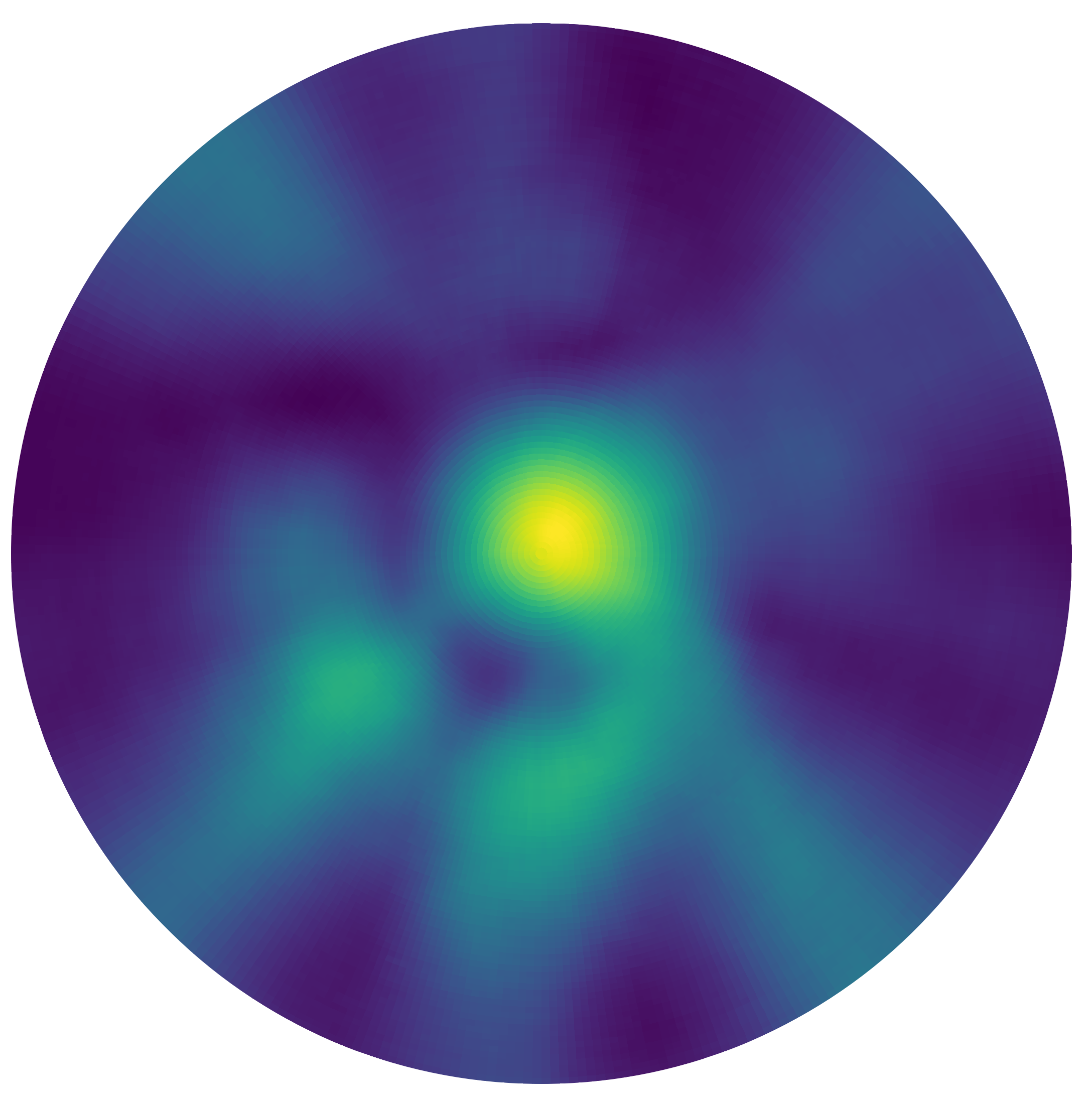} &
    \colstack{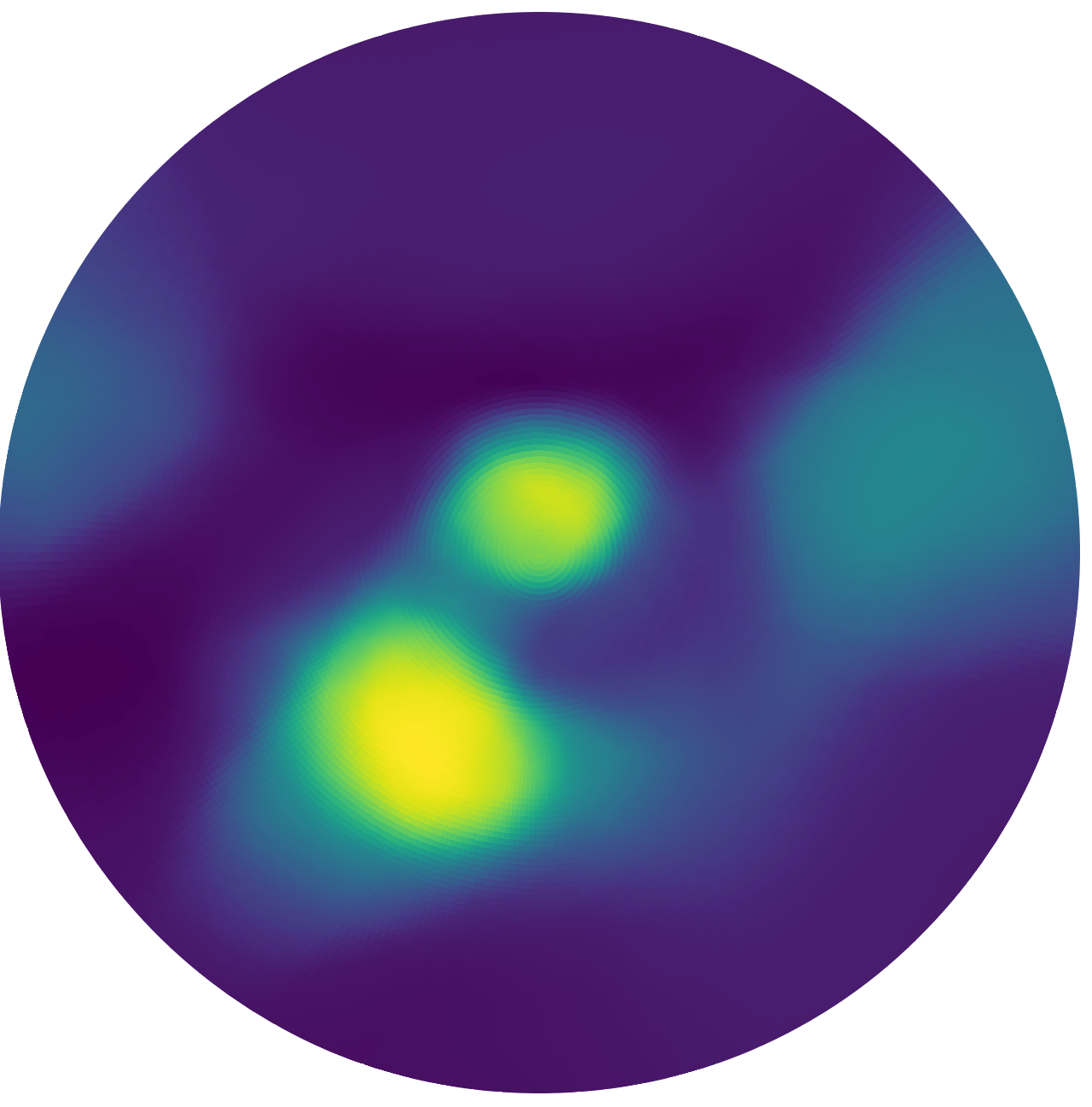}{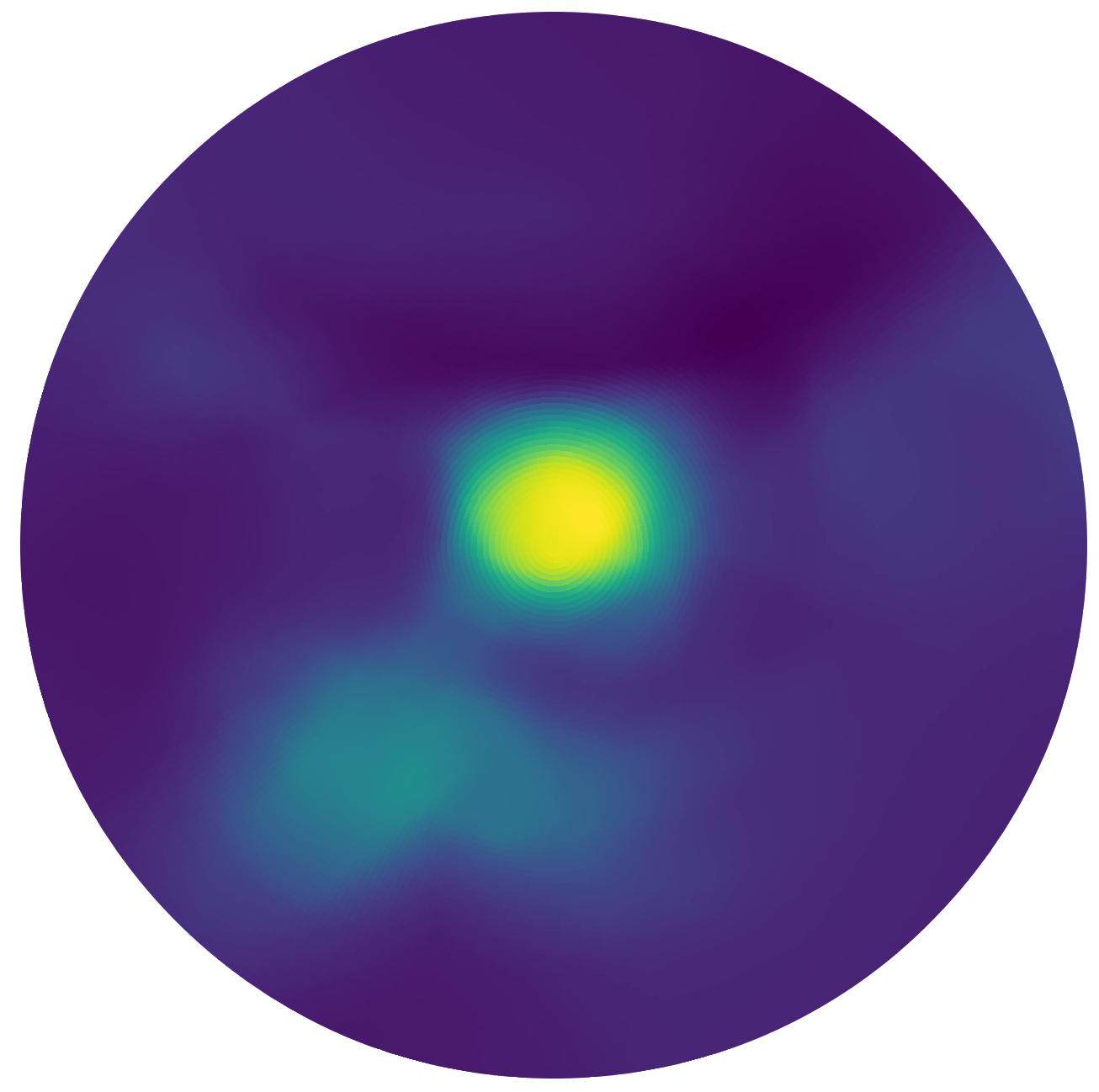}{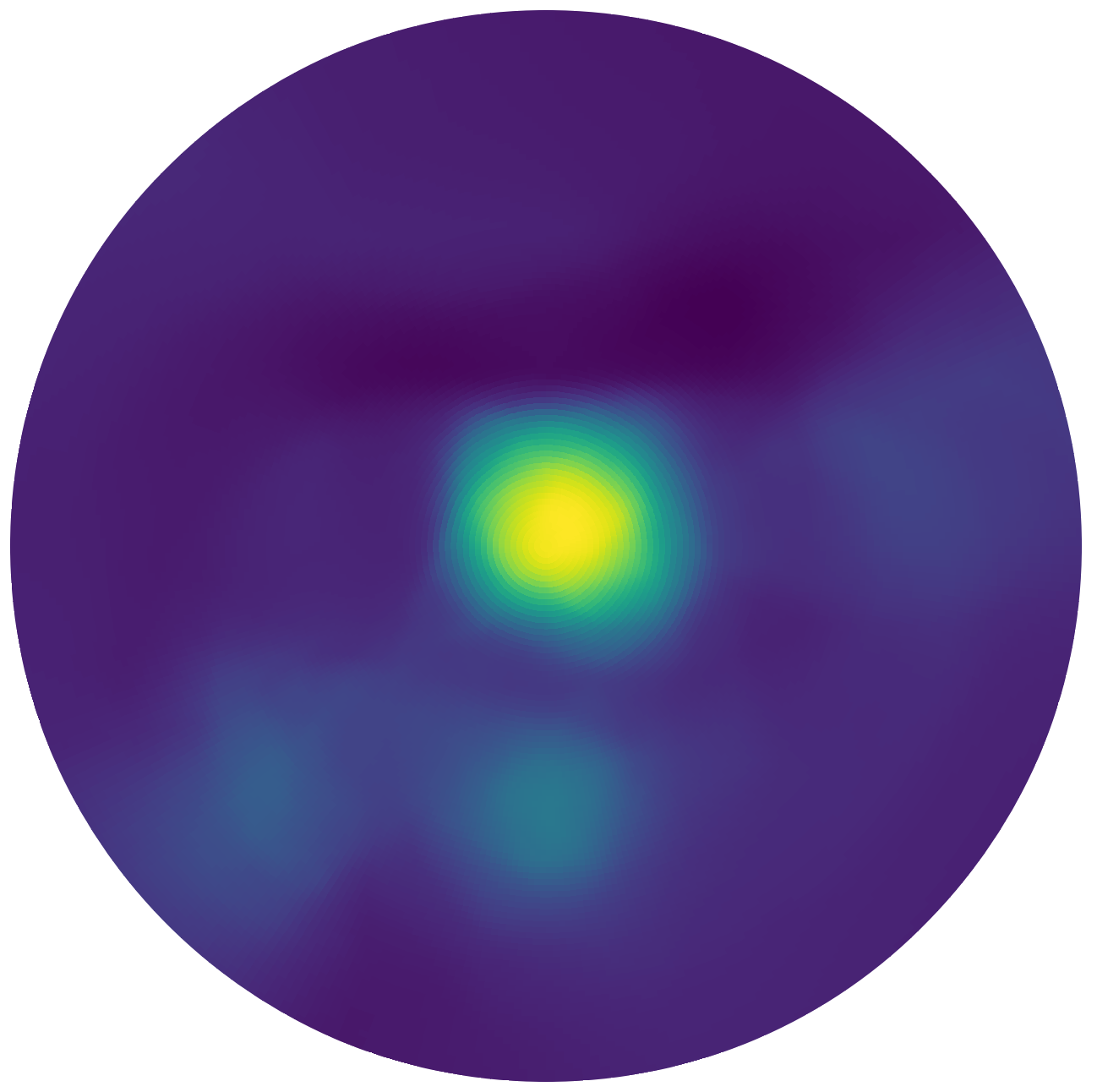} &
    \colstack{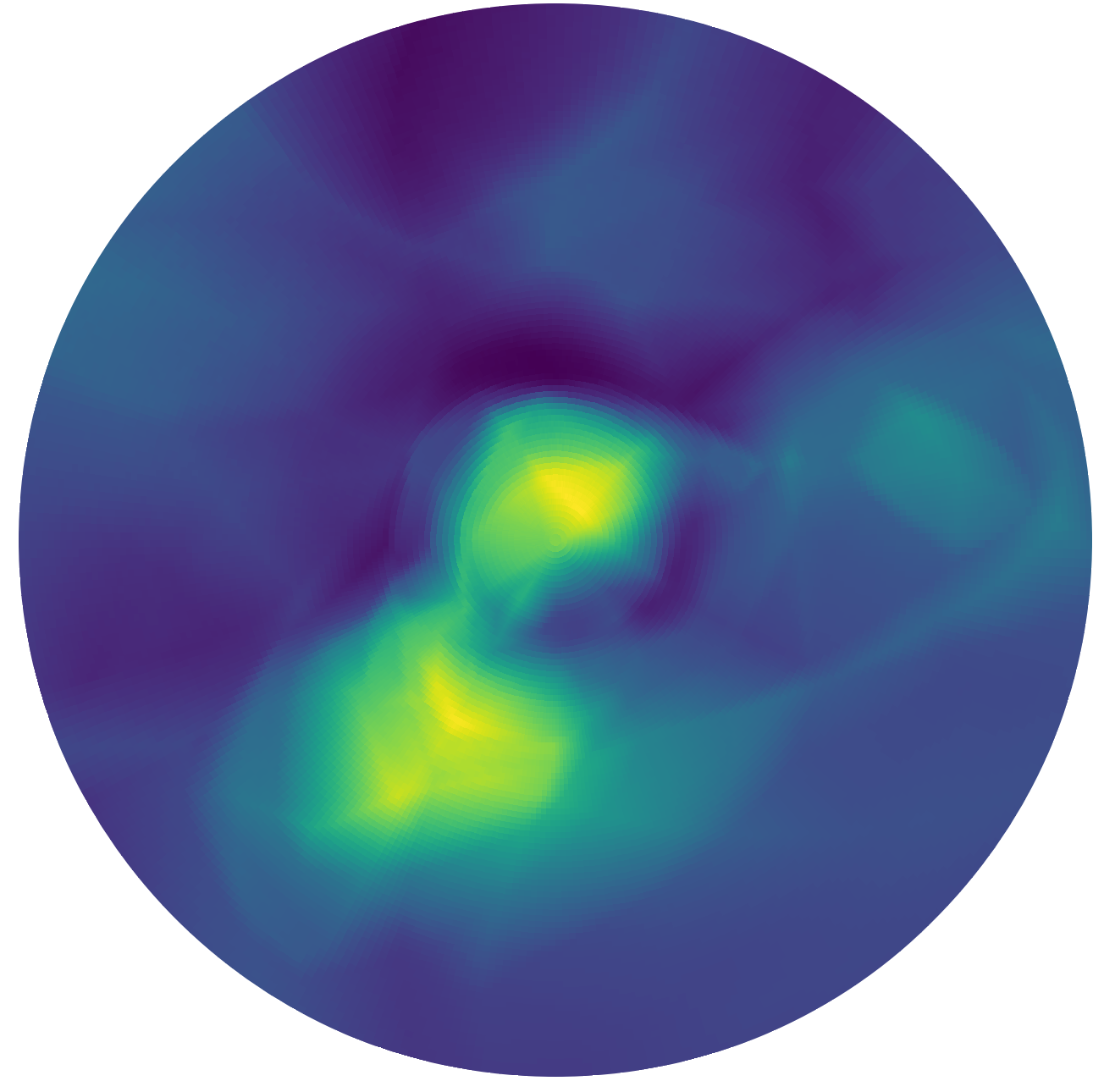}{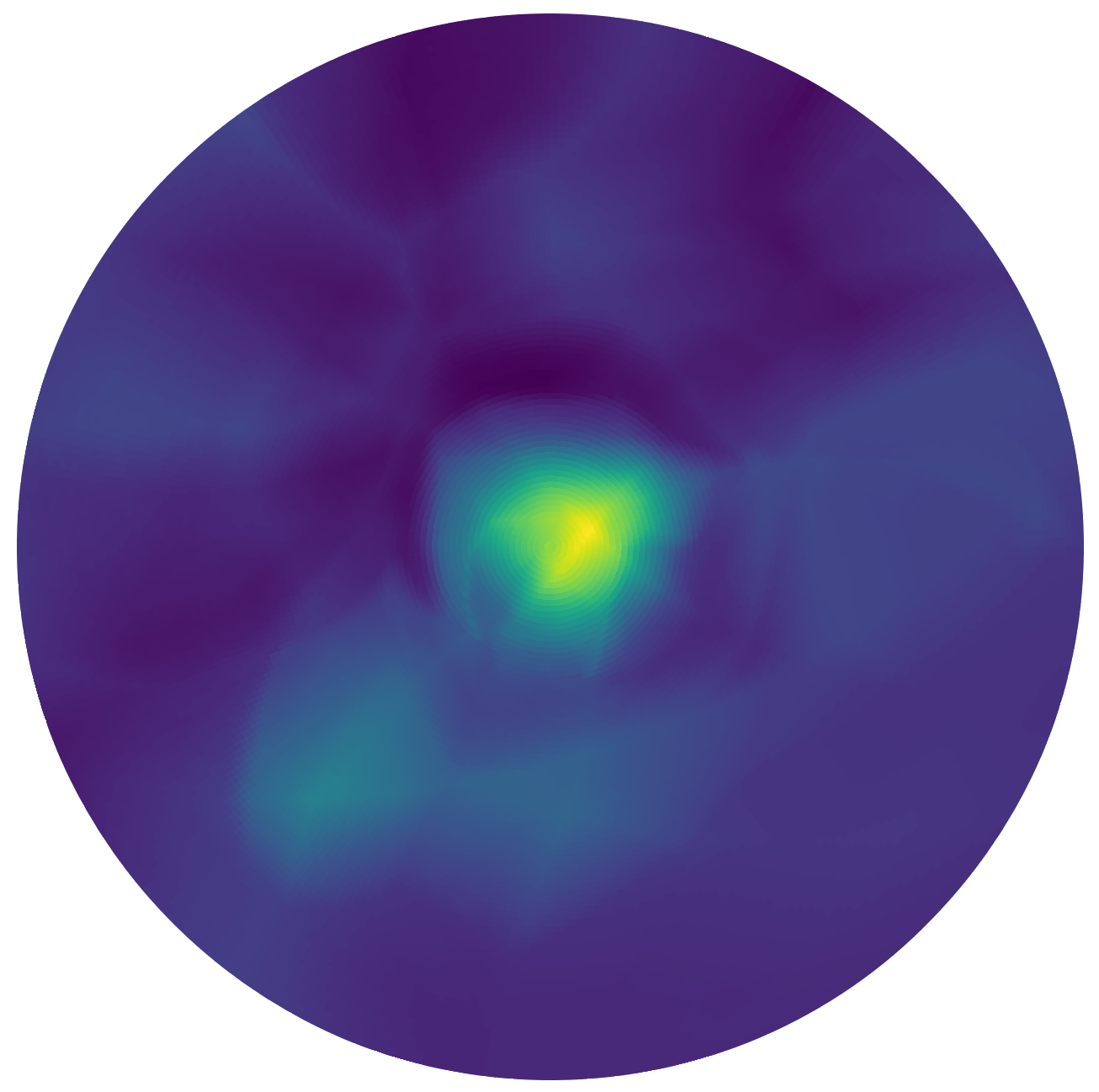}{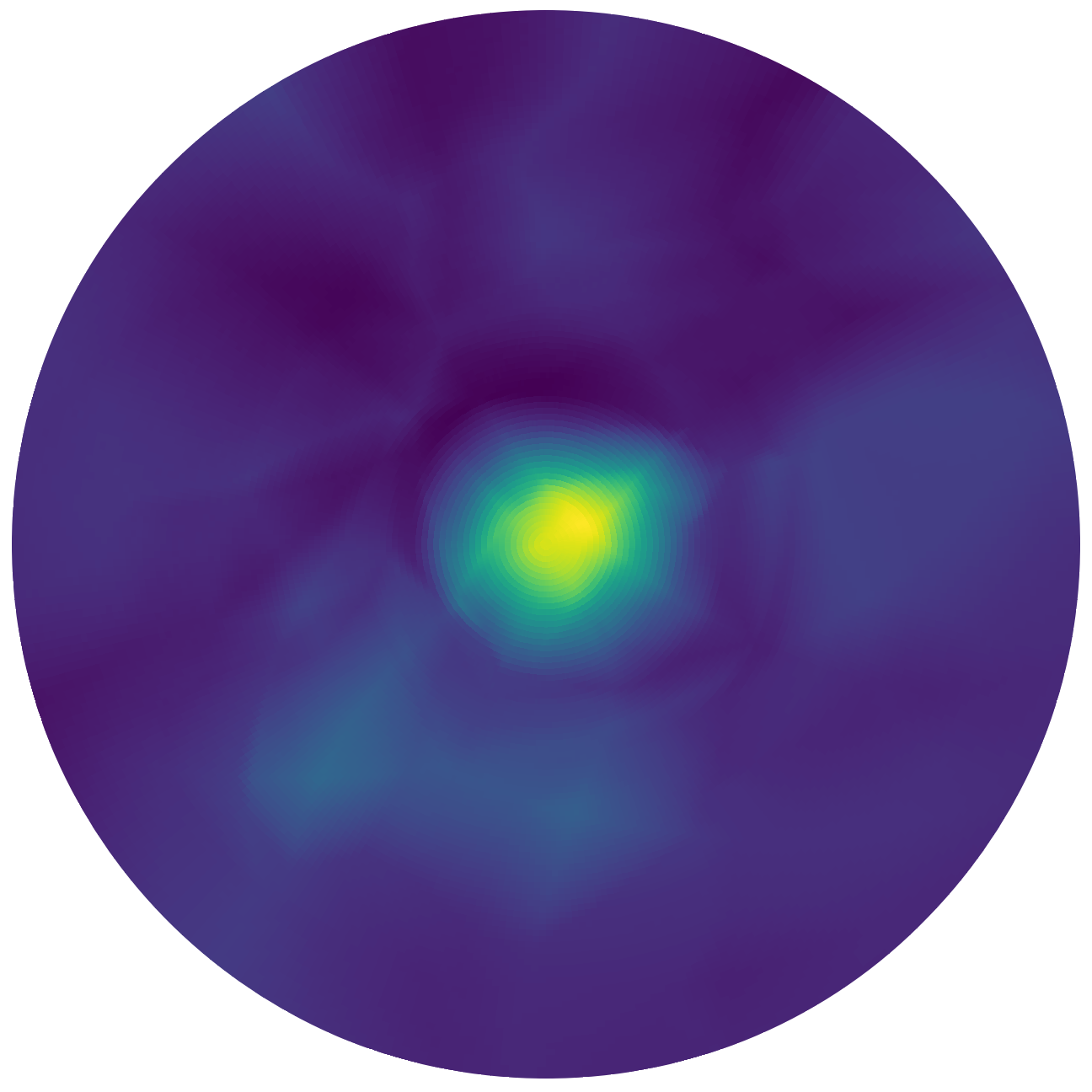} &
    \colstack{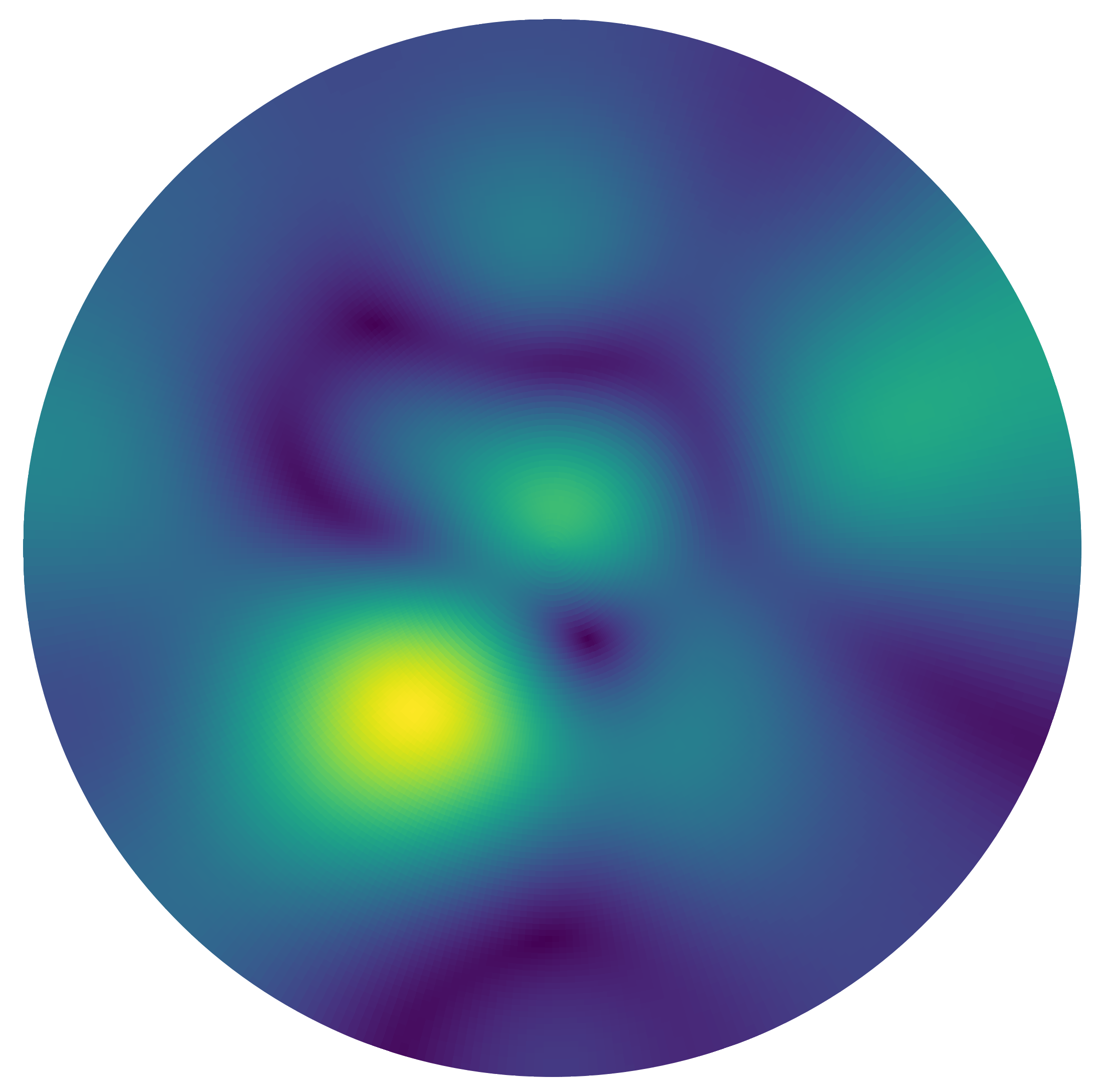}{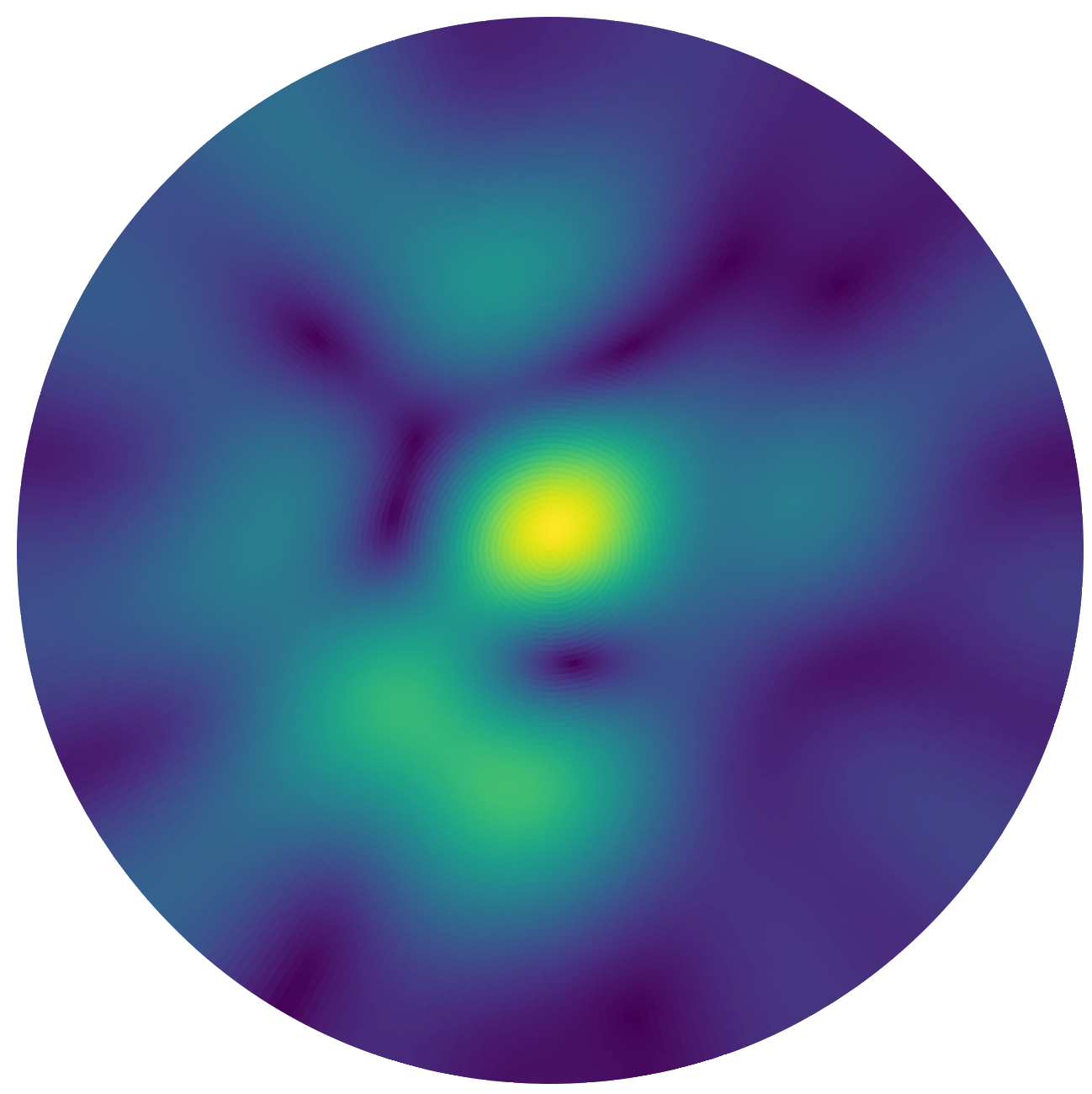}{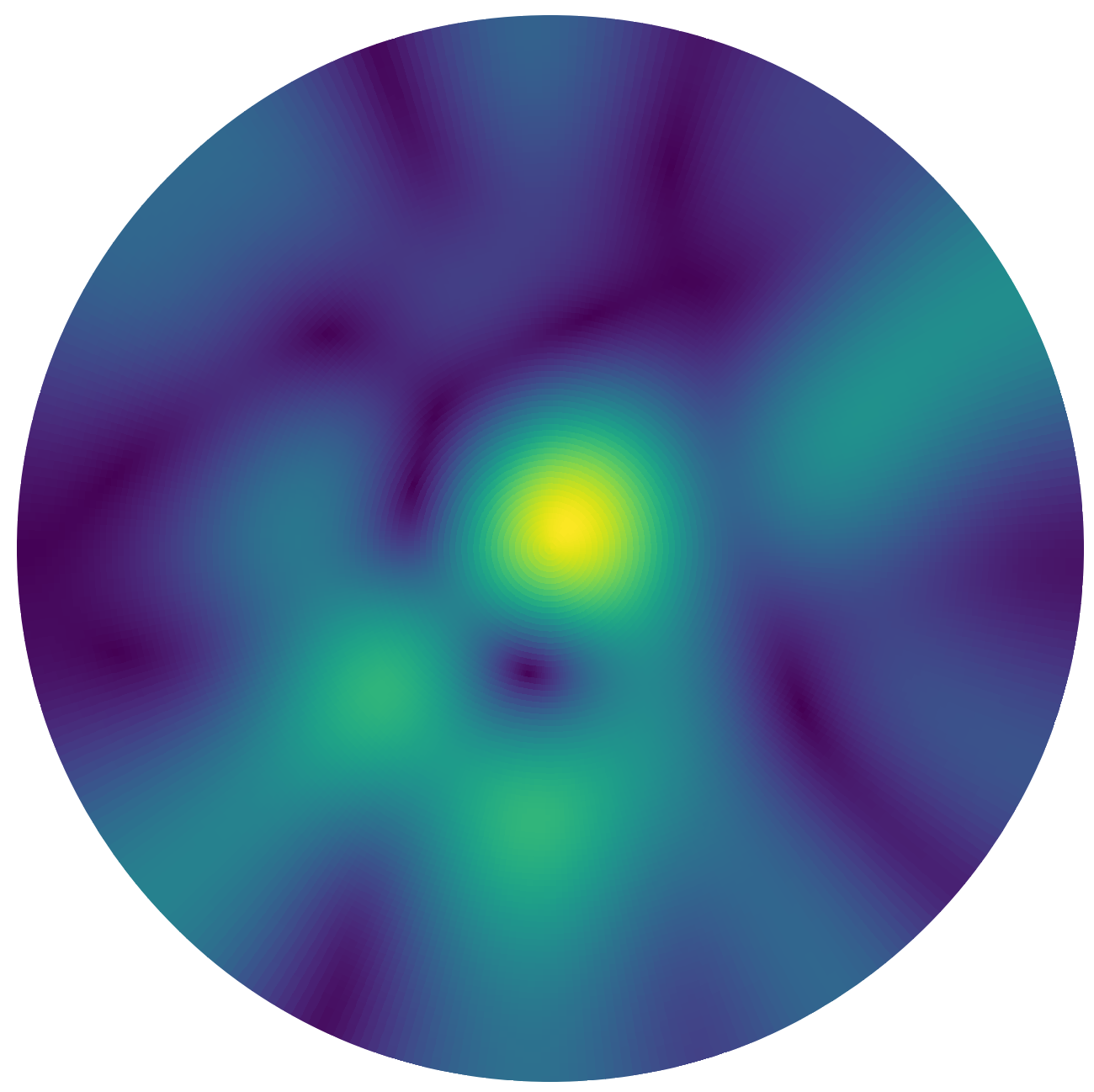}
    \\[-0.5mm]

    \makebox[\colw][c]{\footnotesize (a) Ours} &
    \makebox[\colw][c]{\footnotesize (b) WRF-GS} &
    \makebox[\colw][c]{\footnotesize (c) WRF-GS+} &
    \makebox[\colw][c]{\footnotesize (d) DCGAN} &
    \makebox[\colw][c]{\footnotesize (e) NeRF$^2$} &
    \makebox[\colw][c]{\footnotesize (f) GAI-NeRF} &
    \makebox[\colw][c]{\footnotesize (g) MLP} &
    \makebox[\colw][c]{\footnotesize (h) FIRE} &
    \makebox[\colw][c]{\footnotesize (i) Label}

  \end{tabular}%
  }

  \vspace{-1.5mm}
  \caption{2D spatial spectrum visualizations.}
  \label{fig:2dvisualization}
\end{figure*}

\paragraph{Boundary attenuation regularization.}

The soft mask also defines an implicit air--material interface through its transition band:
\begin{equation}
\mathcal{B}=\left\{i:\left|m(\mathbf{x_i})-0.5\right|<\delta_{\mathrm{b}}\right\},
\end{equation}
where $\delta_{\mathrm{b}} \in (0,0.5)$ is a boundary-band tolerance that controls the thickness of the soft transition region around $m=0.5$. On this learned boundary, we impose an absorption-inspired condition that penalizes strong field responses near wave-blocking material:
\begin{equation}
\mathcal{L}_{\mathrm{BC}}=\frac{1}{|\mathcal{B}|}\sum_{i\in\mathcal{B}}\left|u(\mathbf{x_i})\right|^{2}.
\end{equation}
This term captures the intuition that obstacle interfaces often attenuate or redirect the free-space propagation field rather than support unimpeded wave propagation. It also couples the field and mask predictions: the mask determines where the boundary is located, while the field is constrained to decay along that boundary. Thus, the model can infer obstacle interfaces and their propagation effects from wireless measurements and physics regularization, without requiring explicit boundary labels or material annotations.

\paragraph{Mask regularization.}

The medium mask must remain both interpretable and useful for physics weighting. Without regularization, the model may keep large portions of the scene around $m\approx0.5$, which makes the free-space region, obstacle region, and boundary band ambiguous. We therefore regularize the mask using an entropy term and a sparsity term:
\begin{equation}
\mathcal{L}_{\mathrm{mask}}=\lambda_{\mathrm{ent}}\mathbb{E}\!\left[-m\log m-(1-m)\log(1-m)\right]+\lambda_{\mathrm{spa}}\mathbb{E}[m].
\end{equation}
The entropy weight  $\lambda_\mathrm{ent}$ encourages the mask to approach a near-binary free-space/obstacle partition, while the sparsity term $\lambda_\mathrm{spa}$ discourages the model from explaining propagation loss by marking excessive regions as obstacles. Together, these terms promote a compact and physically meaningful medium decomposition, which stabilizes both the Helmholtz residual in free space and the boundary loss near implicit interfaces. Both terms depend only on the predicted mask and require no external geometry target. The mask is therefore learned without CAD models, floor plans, or manual material annotations.

\subsection{Joint Objective and Gradient Coupling}
The physics-informed objective aggregates the three complementary constraints introduced above,
\begin{equation}
\mathcal{L}_{\mathrm{PINN}}=\lambda_{\mathrm{PDE}}\mathcal{L}_{\mathrm{PDE}}+\lambda_{\mathrm{BC}}\mathcal{L}_{\mathrm{BC}}+\lambda_{\mathrm{mask}}\mathcal{L}_{\mathrm{mask}},
\end{equation}
where $\lambda_{\mathrm{PDE}},\lambda_{\mathrm{BC}}$ and $\lambda_{\mathrm{mask}}$ are the hyperparameters for each physical regularization loss. Then it is combined with the data term into the full training objective
\begin{equation}
\mathcal{L}=\mathcal{L}_{\mathrm{data}}+\lambda_{\mathrm{PINN}}\mathcal{L}_{\mathrm{PINN}}.
\end{equation}
Since the field and mask heads share the backbone with the rendering branch, $\mathcal{L}_{\mathrm{PINN}}$ regularizes the same latent Gaussian representation that produces the rendered observations. The two branches do not exchange forward signals, as the rasterizer does not consume $(u,m)$ and the Helmholtz operator does not consume the rendered output, so they interact only through back-propagation. This makes the physics branch a training-time regularizer rather than an inference-time module, improving the physical consistency of the learned representation while preserving the rendering efficiency of the underlying 3D Gaussian model at test time.

\section{Datasets and Results}

\begin{table*}[t]
\centering
\caption{Performance comparison of PINN-GS against baseline methods across different room configurations and datasets. Lower Median Absolute Error~(MedAE) values and higher Structural Similarity~(SSIM) values indicate better performance.}
\label{tab:performance_comparison}
\setlength{\tabcolsep}{5pt}
\renewcommand{\arraystretch}{0.9}
\resizebox{0.7\linewidth}{!}{
\begin{tabular}{@{}lccccc@{}}
\toprule
\multirow{2}{*}{\textbf{Method}} 
& \multicolumn{2}{c}{\textbf{Room 1 / MedAE (dB) $\downarrow$}} 
& \multicolumn{2}{c}{\textbf{Room 2 / MedAE (dB) $\downarrow$}} 
& \textbf{Other Dataset} \\
\cmidrule(lr){2-3} \cmidrule(lr){4-5} \cmidrule(l){6-6}
& \textbf{2.4 GHz} & \textbf{5.0 GHz} 
& \textbf{2.4 GHz} & \textbf{5.0 GHz} 
& \textbf{Spectrum SSIM $\uparrow$} \\
\midrule
Ray Tracing~\cite{7152831}        & 25.5 & 20.7 & 25.6 & 20.8 & 0.33 \\
MLP                               & 7.3  & 9.3  & 8.2  & 9.9  & 0.71 \\
FIRE~\cite{10.1145/3447993.3483275} & 5.8 & 5.5 & 4.5 & 2.7 & 0.73 \\
DCGAN~\cite{radford2015unsupervised} & 4.0 & 3.4 & 4.2 & 3.0 & 0.56 \\
NeRF$^2$~\cite{nerf2} & 3.6 & 2.9 & 3.9 & 2.0 & 0.78 \\
GAI-NeRF~\cite{shen2026geometricnerf} & 3.2 & 2.1 & 3.1 & 1.8 & 0.84 \\
WRF-GS~\cite{wrfgs}               & 3.3  & 2.4  & 3.4  & 1.9  & 0.82 \\
WRF-GS+~\cite{wrfgs-plus}         & 3.1 & 2.1 & 2.9 & 1.9 & 0.89 \\
\midrule
\textbf{PINN-GS~(ours)}                  & \textbf{3.0} & \textbf{1.9} & \textbf{2.6} & \textbf{1.5} & \textbf{0.92} \\
\bottomrule
\end{tabular}}
\end{table*}

\subsection{Datasets}
We evaluate PINN-GS across two complementary scenarios. First, we use the public RFID spectrum dataset~\cite{nerf2} to examine high-dimensional wireless field reconstruction, where array measurements are converted into front-hemisphere spatial spectra over an azimuth--elevation grid. For this dataset, the model predicts the spectrum magnitude on the azimuth--elevation grid, which is directly compared with the reference spectrum. This setting evaluates whether the model can preserve fine-grained angular structures and recover spatially coherent RF energy distributions. Second, we evaluate on two indoor room datasets~\cite{shen2026geometric3d, shen2026geometricnerf} with dense RSSI measurements under realistic deployment conditions. These rooms contain structural obstacles and occlusions that induce attenuation, blockage, and multipath effects, making them suitable for assessing scalar wireless field reconstruction in practical indoor environments. Specifically, we evaluate a real-world indoor RSSI dataset over an approximately $35\,\mathrm{m}^2$ indoor site using a semi-automated TurtleBot 4 platform. The robot records spatially referenced poses with an RPLIDAR A1M8 and the Nav2+AMCL stack, while RSSI measurements are collected at both 2.4 GHz and 5 GHz using a Raspberry Pi 4 with a dual-band TP-Link Archer T3U AC1300 adapter. An ASUS RT-AC86U router is fixed at $1\,\mathrm{m}$ height as the transmitter. At each waypoint, the platform records the robot pose and five RSSI samples per band. Together, the spectrum dataset and the two-room RSSI datasets cover both directional RF field rendering and site-specific received-signal prediction, allowing us to evaluate PINN-GS under complementary wireless sensing modalities.

\subsection{Baselines}
We compare PINN-GS against a representative set of wireless field reconstruction methods spanning four categories. As a deterministic physical baseline, we use MATLAB's ray-tracing toolbox~\cite{7152831}, which predicts the received signal at each Rx location from a 3D scene model and the Tx position. For learning-based regression, we include a multilayer perceptron (MLP) and FIRE~\cite{10.1145/3447993.3483275}. As a generative baseline, we evaluate a DCGAN~\cite{radford2015unsupervised}. For neural-field methods, we adopt the standard NeRF$^2$~\cite{nerf2}. We also include GAI-NeRF~\cite{shen2026geometricnerf} as a geometric-algebra-informed neural-field baseline for wireless channel prediction. Finally, we benchmark against the state-of-the-art Gaussian-splatting wireless models WRF-GS~\cite{wrfgs} and WRF-GS+~\cite{wrfgs-plus}. All baselines are trained and evaluated under identical data splits for a fair comparison.

\subsection{Results}

Following prior wireless field reconstruction studies~\cite{nerf2,wrfgs,wrfgs-plus,shen2026geometricnerf}, we report point estimates on the held-out test split. Table~\ref{tab:performance_comparison} compares PINN-GS with representative physical, learning-based, neural-field, and Gaussian-splatting baselines. The large error of ray tracing indicates that purely geometry-based propagation modeling is difficult to apply reliably in practical indoor settings without accurate material properties and careful calibration. Learning-based baselines reduce this error, but generic regressors and generative models still struggle to capture the spatial structure of wireless propagation. Neural-field and Gaussian-splatting methods perform substantially better, showing the benefit of continuous scene-specific wireless representations. Building on this direction, PINN-GS achieves the best overall performance across both RSSI and spectrum reconstruction tasks. Compared with the strongest Gaussian baseline, WRF-GS+, PINN-GS reduces the Room~2 RSSI MedAE from $2.9$ to $2.6$ dB at $2.4$ GHz and from $1.9$ to $1.5$ dB at $5.0$ GHz, while also improving the RFID spectrum SSIM from $0.89$ to $0.92$. These gains suggest that scene-aware wave regularization provides complementary guidance beyond data fitting, improving both scalar received-signal prediction and high-dimensional RF spectrum reconstruction.

\begin{table}[t]
  \centering
  \caption{Ablation study of the physics-informed components. Each variant only removes one component. Lower MedAE is better; the best result for each dataset is shown in \textbf{bold}.
  }
  \label{tab:pinn-ablation}

  \setlength{\tabcolsep}{3pt}
  \small
  \resizebox{\columnwidth}{!}{%
    \begin{tabular}{lcccccc}
      \toprule
      Dataset / MedAE (dB)
      & Ours
      & w/o PDE
      & w/o BC
      & w/o Mask
      & w/o Lap.
      & w/o Coup. \\
      \midrule
      \textsc{Room2 / 2.4 GHz}
      & \textbf{2.6}
      & 3.3
      & 3.0
      & 3.6
      & 3.2
      & 3.2 \\
      
      \textsc{Room2 / 5 GHz}
      & \textbf{1.5}
      & 1.8
      & 1.9
      & 1.8
      & 1.6
      & 1.7 \\
      \bottomrule
    \end{tabular}%
  }
\end{table}

Fig.~\ref{fig:2dvisualization} qualitatively illustrates the spectral fidelity gains in Table~\ref{tab:performance_comparison}. Purely data-driven baselines such as MLP and FIRE over-smooth the spectrum, while DCGAN introduces spurious high-energy artifacts, consistent with its low SSIM. Neural-field and Gaussian-splatting baselines better recover the dominant peak, but still blur angular boundaries and miss fine secondary structures. In contrast, PINN-GS more closely matches the reference by preserving both the peak sharpness and surrounding low-energy patterns. This improvement stems from the physics-informed branch, which enforces free-space Helmholtz consistency and boundary-aware attenuation, leading to more angularly coherent RF energy distributions. These qualitative results are consistent with the quantitative gains in Table~\ref{tab:performance_comparison}. The ablation study below further evaluates the contribution of the physics-informed components.

Table~\ref{tab:pinn-ablation} presents our ablation study of the physical components. We remove the Helmholtz residual, boundary regularization, and mask regularization, respectively. The final two variants examine the physical mechanism more directly. \emph{w/o Lap.} removes only the differential term from the wave residual, while \emph{w/o Coup.} retains the full physical objective but blocks its gradients from reaching the shared backbone. Our model achieves the lowest error at both frequencies. Among the loss-term ablations, removing mask regularization has the largest effect at the lower frequency, while removing boundary regularization has the largest effect at the higher frequency. This pattern indicates that the two terms provide complementary guidance across propagation settings.

\begin{table}[t]
\centering
\caption{Collocation resolution relative to the effective wavelength on which the constraint acts. we use a scene-scaled effective wavenumber \(k_{\mathrm{eff}}\). Accordingly, \(\lambda_{\mathrm{eff}}=2\pi/k_{\mathrm{eff}}\).}
\label{tab:collocation}
\small
\begin{tabular}{lcccc}
\toprule
Dataset & Scene extent & $\lambda_\mathrm{eff}$ & Spacing & Samples\,/\,$\lambda_\mathrm{eff}$ \\
\midrule
Room 1 / 2.4\,GHz & 9.49\,m & 2.37\,m & 0.096\,m & 24.7 \\
Room 1 / 5\,GHz & 9.83\,m & 2.46\,m & 0.128\,m & 19.3 \\
Room 2 / 2.4\,GHz & 8.42\,m & 2.10\,m & 0.135\,m & 15.6 \\
Room 2 / 5\,GHz & 8.99\,m & 2.25\,m & 0.108\,m & 20.9 \\
Spectrum (RFID) & 4.81\,m & 1.20\,m & 0.080\,m & 15.0 \\
\bottomrule
\end{tabular}
\end{table}

Table~\ref{tab:collocation} relates the collocation resolution to the spatial scale of the physical constraint. Scene extent is the diagonal span of the Gaussian scene and summarizes the size of the represented region. The effective wavelength is derived from this extent and defines the scale at which the wave constraint acts. Spacing is the median nearest-neighbor distance among the sampled Gaussian centers. The final column reports the effective wavelength divided by this spacing, so larger values indicate finer resolution relative to the wave scale. Across all datasets, each effective wavelength contains many collocation samples, which provide sufficient spatial resolution to apply the wave constraint at its intended scale.

\begin{figure}[htbp]
\centering
\includegraphics[width=0.86\linewidth]{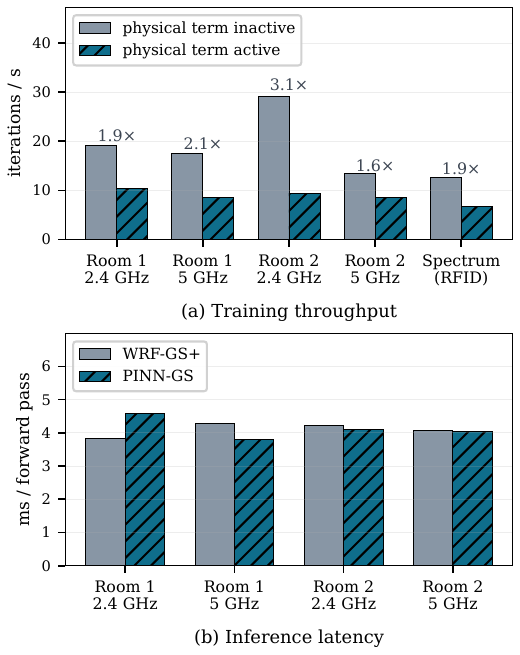}
\caption{Training and inference cost comparison.}
\label{fig:cost}
\end{figure}

Fig.~\ref{fig:cost} separates the training overhead of physical supervision from the inference cost of the final renderer. Fig.~\ref{fig:cost}(a) compares training throughput with the physical term disabled and enabled. Enabling the physical loss reduces throughput because computing the PDE residual requires field derivatives at the collocation points. In contrast, Fig.~\ref{fig:cost}(b) shows similar inference latency for PINN-GS and WRF-GS+ across the evaluated rooms. The field and mask heads are used only during training and are skipped at inference.

\section{Conclusion}
In this paper, we presented PINN-GS, the first physics-informed 3D Gaussian Splatting framework for wireless radiation field reconstruction. %This work is to make physical consistency an integral property of the Gaussian scene representation, rather than an auxiliary constraint separated from the rendering process. 
By coupling differentiable wireless rendering with scene-aware wave regularization, PINN-GS encourages the reconstructed field to respect the spatial structure of electromagnetic propagation while maintaining the efficiency and adaptivity of Gaussian primitives. The learned wireless wave field and soft propagation mask further provide a compact mechanism for aligning field reconstruction with the underlying propagation behavior of the environment. Across RSSI and RFID spectrum benchmarks, our method achieves consistent improvements in signal prediction accuracy and spectral fidelity, demonstrating that physics-aware Gaussian representations can produce more faithful wireless fields. %These results suggest a promising direction for unifying neural scene representations and electromagnetic priors in wireless sensing and communication systems.

\begin{acks}
This work is supported in part by the NSF (CNS-2415209, IIS-2543148, CNS-2317190, IIS-2306791, CNS-2319343).
\end{acks}

{
    \small
    \bibliographystyle{ACM-Reference-Format}
    \bibliography{ref}
}

\appendix
\section{Physical Behaviour of the Learned Field}
\begin{figure}[htbp]
  \centering
  \includegraphics[width=0.95\columnwidth]{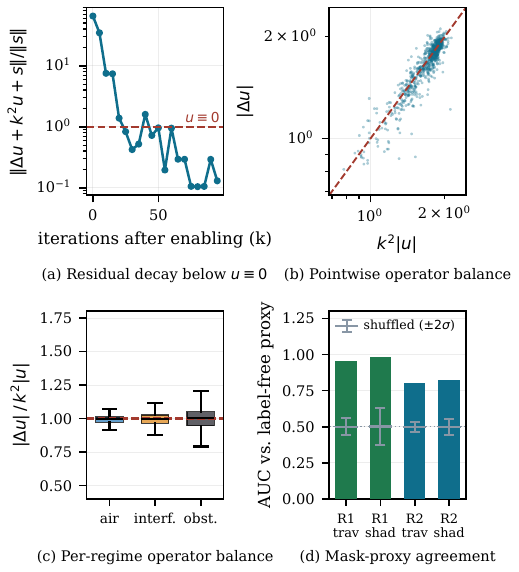}
  \caption{Behaviour of the training-time wave constraint, evaluated at the
  collocation points.}
  \label{fig:physics-audit}
\end{figure}

Fig.~\ref{fig:physics-audit} evaluates the training-time latent field at the collocation points where the wave constraint is applied. It addresses four questions: whether the field is non-trivial, whether both operator terms remain active, whether their balance holds across the learned mask regimes, and whether the mask aligns with measurement-derived scene cues. The wave residual is normalized by the source strength, which gives the trivial field $u \equiv 0$ a fixed analytic reference independent of the source amplitude. 

Panel~(a) shows that the residual moves from above this reference to values below it during training. The learned field therefore attains a source-normalized residual below the zero-field reference. Panel~(b) compares the Laplacian and wavenumber terms at the collocation points. Their balanced magnitudes, together with the residual trend in panel~(a), show that both terms remain active and link the residual reduction to operator balance rather than amplitude scaling. Panel~(c) shows the same balance across the air, interface, and obstacle-like regimes, indicating that mask-based weighting preserves this operator behavior throughout the learned scene partition. Panel~(d) evaluates the learned mask against two label-free proxies derived from the wireless measurements. The observed agreement is clearly separated from the shuffled null distributions in both rooms. This shows that the learned partition aligns with scene-related cues in the measurements without requiring geometric annotations.

The field and rendering heads both consume the same transmitter-conditioned latent feature. Because the physical loss backpropagates through the shared backbone, it directly regularizes the representation used for rendering. Together, the panels show that the physical branch learns a non-trivial field, maintains Helmholtz operator balance across the mask regimes, and weights the constraint with a scene-aware mask.

\begin{figure}[htbp]
\centering
\includegraphics[width=\linewidth]{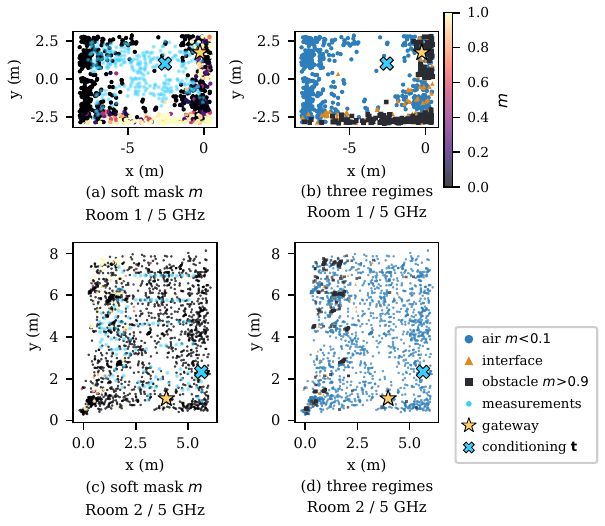}
\caption{Medium masks learned on two rooms, each conditioned on the marked
  transmitter position $\mathbf{t}$.}
\label{figure_mask_measured}
\end{figure}

\begin{figure}[htbp]
  \centering
  \includegraphics[width=0.9\columnwidth]{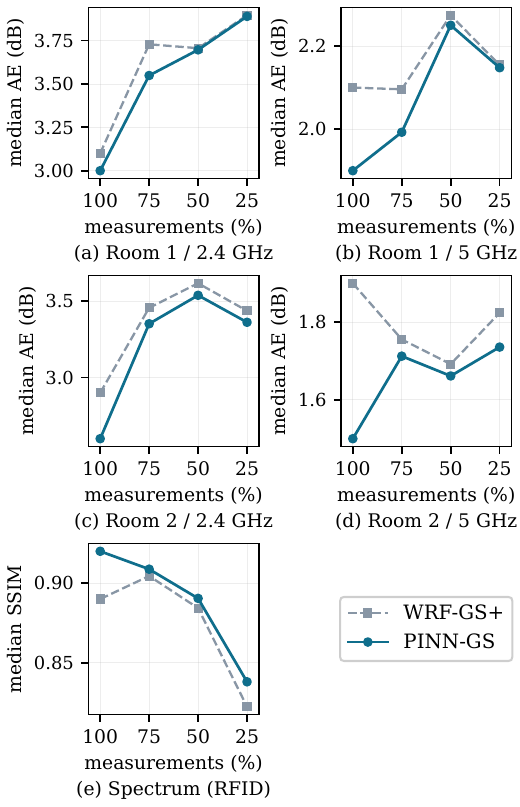}
  \caption{Reconstruction accuracy as measurements are reduced. Both methods
  train on the same subsets; values at 100\% are those in Table~\ref{tab:performance_comparison}.}
  \label{fig:sparsity}
\end{figure}

\section{The Learned Medium Mask}
\label{app:mask}

Fig.~\ref{figure_mask} illustrates the conceptual partition defined by the soft mask. Fig.~\ref{figure_mask_measured} shows the corresponding transmitter-aware masks learned from wireless measurements in two rooms. Panels~(a) and~(c) show the continuous mask over each Gaussian scene. Low-mask values form broad connected regions, while high-mask responses concentrate in localized areas and near scene boundaries. Intermediate values create smooth transitions, allowing the physical weighting to vary continuously across the scene.

Panels~(b) and~(d) map each continuous mask to the air, interface, and obstacle-like regimes used by the physical objective. The resulting partitions contain connected air-like regions, compact obstacle-like structures, and transition regions at their interfaces. These regimes determine where the free-space residual is emphasized, where its weighting is adapted to the learned medium, and where boundary regularization is applied.

The two rooms produce distinct masks, and each mask varies with the transmitter position that conditions it. This variation shows that the mask adapts to both the scene representation and the transmitter configuration rather than acting as a fixed spatial prior. Training requires no CAD model, floor plan, mesh, or material label. The learned masks are also consistent with the label-free, measurement-derived proxies reported in Figure~\ref{fig:physics-audit}. Together, the spatial visualizations and proxy tests support the use of the mask as a structured, scene-aware signal that reflects the propagation-relevant geometry of each room and directs the physical constraint to the regions where it applies.

\section{Reconstruction under Sparse Measurements}
\label{app:sparsity}

Figure~\ref{fig:sparsity} evaluates how each method degrades as measurements are removed. Both lose accuracy as the density drops, which is expected when the field is constrained by fewer observations; the question is whether the physical prior still helps once data become scarce. PINN-GS keeps its advantage on all five datasets down to half the measurements, and at one quarter it remains ahead on the 5 GHz rooms and on the spectrum benchmark. On the two 2.4 GHz rooms the quarter-density setting leaves only a few hundred training samples, and the difference between the two methods there falls within the spread produced by the choice of subset. The behaviour is consistent with the role of the physical term: where measurements no longer determine the field, the wave constraint continues to restrict it. The curves stop at one quarter because below this density the comparison would reflect the particular subset drawn rather than the effect of the prior.

\end{document}